\documentclass[lettersize,journal]{IEEEtran}

\IEEEoverridecommandlockouts

\usepackage{amsmath,amssymb,amsfonts}
\usepackage{bm}
\usepackage{array}
\usepackage[caption=false,font=normalsize,labelfont=sf,textfont=sf]{subfig}
\usepackage{textcomp}
\usepackage{stfloats}
\usepackage{url}
\usepackage{verbatim}
\usepackage{graphicx}
\def\BibTeX{{\rm B\kern-.05em{\sc i\kern-.025em b}\kern-.08em
    T\kern-.1667em\lower.7ex\hbox{E}\kern-.125emX}}
\usepackage{balance}

\usepackage{algorithmic}
\usepackage{graphicx}
\usepackage{cite}
\usepackage{xcolor}
\usepackage{enumitem}
\usepackage{bm} 
\usepackage{algorithmic}
\usepackage{algorithm}
\usepackage{mathrsfs}
\usepackage{makecell}  
\usepackage{amsmath}   
\usepackage[labelformat=simple]{caption}
\usepackage{stfloats}
\newcommand{\figureref}[1]{Fig.~\ref{#1}}
\def\BibTeX{{\rm B\kern-.05em{\sc i\kern-.025em b}\kern-.08em
    T\kern-.1667em\lower.7ex\hbox{E}\kern-.125emX}}

\begin{document}

\title{Low-Complexity Hybrid Precoding for Cell-Free Massive MU-MIMO ISAC Systems\\
\thanks{
Part of this paper was presented at the 2025 IEEE International Conference on Communications (ICC)\cite{b23}. 

Jun Zhu, Yin Xu, Aimin Tang, Ruomeng Wang, Dazhi He, Yunfeng Guan and Wenjun Zhang are from Shanghai Jiao Tong University, Shanghai, 200240, China. The corresponding author is Yin Xu (e-mail: xuyin@sjtu.edu.cn).
}
}

\author{
    \IEEEauthorblockN{Jun Zhu, Yin Xu, \textit{Senior Member, IEEE}, Aimin Tang, \textit{Senior Member, IEEE}, Ruomeng Wang,  \\Dazhi He, \textit{Senior Member, IEEE},  Yunfeng Guan, Wenjun Zhang, \textit{Fellow, IEEE}\\  }
}
\maketitle

\begin{abstract}
Integrated sensing and communication (ISAC) in cell-free (CF) massive multi-user multiple-input multiple-output (MU-MIMO) system is a promising architecture for high-rate communications and high-accuracy multi-target sensing. However, centralized coordination among distributed access points (APs) incurs substantial fronthaul overhead and computation complexity. This paper proposes a low-complexity hybrid precoding framework for CF massive MU-MIMO ISAC systems with partially-connected architectures at the APs. By applying hybrid architecture at the APs, the proposed framework converts the original high-dimensional channel information into a low-dimensional effective channel, enabling digital precoding over the compressed channel domain and thereby substantially reducing both fronthaul overhead and baseband computational complexity. We formulate the joint hybrid precoding design as an ergodic sum-rate (ESR) maximization problem with position error bound (PEB) constraints to ensure multi-target sensing accuracy. An efficient alternating optimization (AO)-based solver is then developed, where the PEB constraint is reformulated into tractable convex constraints, while the digital-domain optimization is carried out over the reduced-dimensional effective channel and the analog precoding is refined on the constant-modulus manifold. For dynamic user topology, we further propose  multi-branch (MB) rate-splitting (RS) minimum mean-square-error Tomlinson-Harashima precoding (MMSE-THP) update algorithm that combines multi-branch ordering with recursive MMSE-THP matrix updates, enabling common and private digital precodings to be refreshed without repeated full matrix recomputation. Simulation results demonstrate that the proposed scheme achieves high ESR and accurate multi-target sensing while reducing computational complexity by 87.02\% compared with conventional baselines.

\end{abstract}


\section{Introduction}
\IEEEPARstart{I}{ntegrated} sensing and communication (ISAC) emerges as a cornerstone technology for sixth-generation (6G) wireless systems, enabling the spectrum resources to alleviate severe frequency scarcity \cite{9737357, 10663787}. In ISAC scenarios, the systems are required not only to support high-performance communication of massive amounts of data, but also to meet the high-precision sensing requirements for targets\cite{11520347, 11421642}. To further boost spectral efficiency (SE) and sensing accuracy, the integration of massive multi-user multiple-input multiple-output (MU-MIMO) into ISAC frameworks has become a pivotal research frontier \cite{11162068,11143948,9999226}. 

A particularly compelling evolution is the cell-free (CF) massive MU-MIMO ISAC architecture\cite{10516289, 11422333}. Unlike traditional cellular setups, the CF paradigm utilizes a central processing unit (CPU) to coordinate geographically distributed access points (APs) via fronthaul links\cite{11271191,10960374}, offering unprecedented multi-angle observations and spatial diversity, which are crucial for resolving target ambiguities and providing uniform communication coverage \cite{11192472,11345189}. In CF massive MU-MIMO ISAC systems, centralized precoding at the CPU enables distributed APs to collaboratively manage multi-user interference and shape sensing beams, making it a key design tool for balancing communication throughput and target-sensing accuracy, with existing strategies being classified into communication-centric and sensing-centric paradigms~\cite{11161262}. Specifically, communication-centric designs optimize the precoding for transmission performance, typically measured by the sum-rate or SE, while incorporating sensing requirements as constraints ~\cite{10978071 }. In contrast, sensing-centric designs optimize the precoding for radar performance, typically measured by target estimation accuracy~\cite{10570706,10982445}. To mitigate the impact of imperfect channel state information (CSI), rate-splitting (RS) schemes are introduced as an advanced technique to enhance the robustness of these precoding strategies within CF massive MU-MIMO frameworks\cite{b26,b29,260601714}. However, incorporating RS into the considered systems further increases the computational burden of centralized precoding design, since both common and private precoding must be jointly optimized across distributed APs.~\cite{9838980,7900388}. Moreover, the continuous exchange of high-dimensional CSI imposes severe overhead on the fronthaul overhead\cite{8891922,9799777}. Consequently, despite the robustness gains of RS, developing an RS-based precoding framework with low computational complexity and reduced fronthaul overhead remains a critical challenge in CF massive MU-MIMO ISAC systems.

In CF massive MU-MIMO ISAC systems, to address these challenges, multiple studies have been carried out. In \cite{10094043},  fronthaul overhead is optimized with learning-based method for practical deployment. In \cite{10879511}, a joint precoding and fronthaul compression scheme is proposed. By leveraging a deep-unfolding neural network based on iterative gradient descent, it effectively mitigates fronthaul overhead while maintaining system performance. The compression of linear precoding algorithms is investigated in \cite{10437610} to satisfy the downlink requirements of wireless sequential fronthaul topologies, which facilitates a substantial reduction in fronthaul transmission overhead and a simultaneous improvement in SE. The study in \cite{10879353} explores the compression of linear precoding algorithms tailored for wireless sequential fronthaul topologies, achieving a substantial reduction in signaling overhead while concurrently enhancing SE. Nevertheless, these approaches still rely on the acquisition or processing of high-dimensional CSI, which remains a fundamental bottleneck in large-scale antenna regimes. Moreover, the computational burden of centralized precoding at the CPU is often left insufficiently addressed.

Beyond fronthaul overhead, computational complexity represents another critical bottleneck in CF massive MU-MIMO ISAC systems, motivating the development of low-complexity precoding. Specifically, low-complexity linear precoding schemes are proposed in \cite{9868348, 10978339, 11288092} to simplify the precoding design and reduce selection complexity. From a resource allocation perspective, the authors in \cite{10494224, 10207026, 11432510,11215654} circumvent high iterative complexity by optimizing the downlink power allocation. Additionally, by exploiting the inherent sparsity of the system, a joint precoding and dedicated AP selection algorithm is designed in \cite{11082576, 9148948}. Despite these advances, existing low-complexity designs have not fully addressed the fronthaul signaling bottleneck caused by high-dimensional CSI exchange. Therefore, developing a unified precoding framework that preserves communication and sensing performance while jointly reducing fronthaul  overhead and computational complexity remains a critical challenge.

Consequently, it is imperative to investigate the precoding frameworks that mitigate both fronthaul and computational complexities while strictly preserving the inherent performance gains of CF massive MU-MIMO ISAC systems. To this end, we propose adopting hybrid architecture at the APs to directly perform dimensionality reduction. Its core advantage lies in the ability to completely decouple the fronthaul transmission overhead from the massive number of transmit antennas, making the overhead depend solely on the limited number of RF chains, thereby significantly alleviating the baseband computational dimensionality at the CPU with a low complexity. To improve communication robustness under imperfect CSI, we incorporate RS into the proposed hybrid precoding framework. However, RS introduces an additional common stream and requires the joint design of common and private digital precoding, which increases the computational burden at the CPU. This burden becomes more pronounced in dynamic user-topology scenarios, where changes in the active user set would otherwise require repeated recomputation of the RS-THP precoding matrices. To address this issue, we develop a low-complexity  update algorithm that reuses previously computed matrix factors to efficiently refresh the digital precoding as the user topology changes.

Furthermore, distinct from prior works primarily limited to single base station\cite{9868348} or single-target detection\cite{11091535}, this paper aims to fully leverage the high spatial diversity of the CF architecture for multi-target environments. Therefore, this paper employs a partially-connected architecture to reduce transmission and computational complexity. However, this approach inherently degrades communication and sensing performance. To address this critical trade-off, we adopt the position
error bound (PEB) to strictly constrain the high-precision sensing performance, based on which a hybrid precoding scheme is designed to maximize the ergodic sum rate (ESR).  Considering that the hybrid architecture at the APs and the digital precoding at the CPU are intricately coupled under non-convex sensing constraints, this paper adopts an alternating optimization (AO) framework to optimize them alternately, ensuring the scheme is  mathematically tractable.
The main contributions of this paper are summarized as follows:
\begin{itemize}
\item First, we formulate a joint design problem to maximize the ESR subject to a PEB constraint, thereby guaranteeing sensing performance. To tackle the resulting non-convexity of this joint design, we develop AO framework that integrates a convex approximation for the PEB constraint, optimal digital precoding, and manifold-based analog precoding. Furthermore, we construct a low-dimensional fronthaul transmission matrix by actively projecting the high-dimensional CSI onto the analog precoding subspace. By leveraging a partially-connected architecture, the proposed scheme significantly mitigates the fronthaul overhead, thereby facilitating a highly efficient transmission framework.
\item Secondly, we propose a novel hybrid precoding scheme designed for CF massive MU-MIMO ISAC systems, which integrates minimum mean square error Tomlinson-Harashima precoding (MMSE-THP) with multi-branch (MB) RS. This framework significantly enhances both SE and sensing accuracy. Furthermore, to address the computational complexity with fluctuating user numbers, an update precoding algorithm is proposed to reduce the computational complexity of the hybrid precoding process while maintaining robust performance.

\item Finally, comprehensive simulation results are provided to evaluate the system in terms of computational complexity, ESR, and beampattern. Simulation results demonstrate that the proposed algorithm reduces computational complexity by 87.02\%, without compromising communication and sensing performance.
\end{itemize}

The rest of this paper is structured as follows. In Section \ref{II}, the communication and sensing models are established, and the corresponding optimization problem is defined. The proposed algorithms are introduced in Section \ref{III}, and the fronthaul overhead is also analyzed. In Section \ref{IV}, an update algorithm is proposed to reduce computational complexity. To validate the effectiveness of these algorithms, comprehensive simulation results are presented in Section \ref{V}. Finally, this paper is concluded in Section \ref{VI}.
 
Notation$:$ Bold uppercase letters denote matrices and bold lowercase letters denote vectors. For a matrix $\textbf H^{\text{H}}$, $\textbf H^{\text{T}}$, $\textbf H^{-1}$,  denote its conjugate transpose, transpose, inverse,  respectively. diag($\textbf H_1, ..., \textbf H_K$) denotes a block diagonal matrix with $\textbf H_1, ..., \textbf H_K$ being its diagonal blocks. The space of $\text{M}\times \text{N}$ complex matrices is expressed as $\mathbb{C}^{\text{M}\times \text{N}}$. Expectations are expressed as $\mathbb{E}[\textbf H]$. $\|\cdot \| _{\text{F}} ^ 2$ represents the square of the norm.

\section{SYSTEM MODEL}
\label{II}

\begin{figure}[t]
\centering
\resizebox{0.35\textwidth}{!}{\includegraphics[]{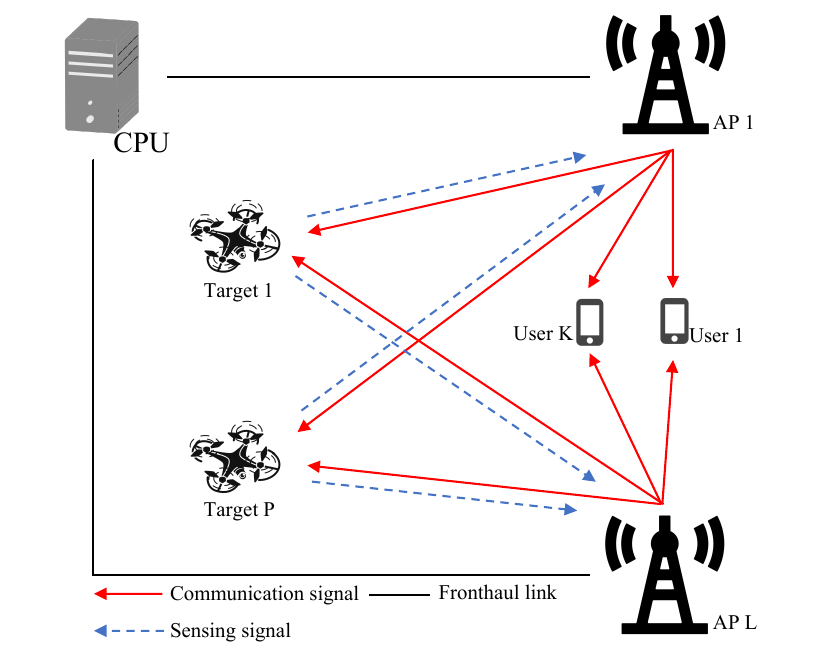}}
\caption{Architecture of the CF massive MU-MIMO ISAC system with hybrid beamforming.}
\label{fig:1}
\end{figure}

\begin{figure}[t]
\centering
\resizebox{0.3\textwidth}{!}{\includegraphics[]{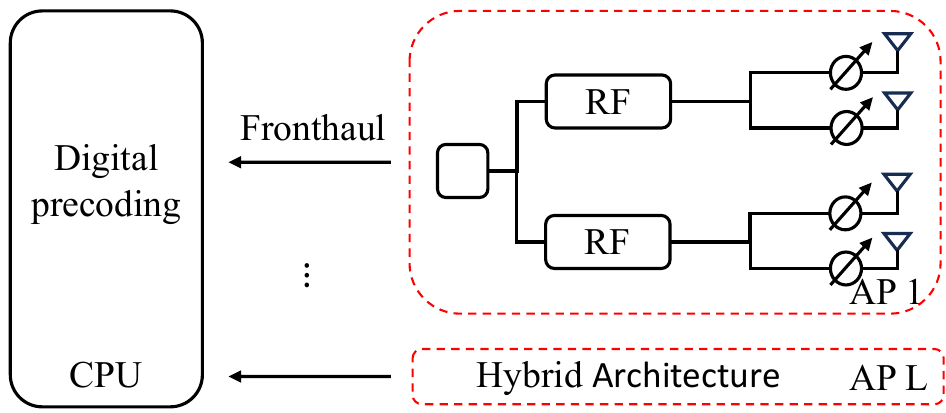}}
\caption{Diagram of CF massive MU-MIMO ISAC system with the partially-connected hybrid architecture.}
\label{fig:2}
\end{figure}

\begin{figure*}[!b]
	\centering
	\vspace*{8pt}
	\hrulefill
	\vspace*{8pt}
	\begin{align}
    \textbf{y}_k &= \sum_{l=1}^L \textbf{H}_k^l \textbf{x}^l + \textbf{n}_k \nonumber = \underbrace{ \sum_{l=1}^L \textbf{H}_k^l \textbf{F}_{\textnormal{RF}}^l \textbf{F}_{\textnormal{BB,c}}^l s_{\textnormal{c}} }_{\textbf{S}_{\textnormal{c},k}} + \nonumber \quad\, \underbrace{ \underbrace{ \sum_{l=1}^L \textbf{H}_k^l \textbf{F}_{\textnormal{RF}}^l \textbf{F}_{\textnormal{BB,p}k}^l \textbf{s}_k }_{ \textbf{S}_{\textnormal{p},k} } + \underbrace{ \sum_{j\neq k}^{K}\sum_{l=1}^{L} \textbf{H}_k^l \textbf{F}_{\textnormal{RF}}^l \textbf{F}_{\textnormal{BB,p}j}^l \textbf{s}_j + \textbf{n}_k }_{ \textbf{I}_{\textnormal{p},k} } }_{\textbf{I}_{\textnormal{c},k}}.  
    \label{eq:2}\tag{2}
	\end{align}
\end{figure*}

As illustrated in Fig. \ref{fig:1}, we consider the downlink of a CF massive MU-MIMO ISAC system. The architecture comprises $L$ APs, where the $l$-th AP is denoted as $\text{AP}~ l$ for $l \in \mathcal{L}$, $ \mathcal{L}\triangleq\{1, 2, \dots, L\}$. These distributed APs are connected to a CPU via fronthaul links. Under the coordination of the CPU, all APs cooperatively serve $K$ multi-antenna communication users and sense $P$ targets.

Regarding the antenna architecture, a partially-connected hybrid structure is adopted at each AP to significantly alleviate computational complexity, as illustrated in Fig. \ref{fig:2}. Specifically, this architecture configures each AP with $N_{\mathrm{l}}$ transmit antennas driven by a reduced number of RF chains $N_{\mathrm{RF}}$, where $N_{\mathrm{RF}} \leq N_{\mathrm{l}}$. Specifically, the $N_{\textnormal{l}}$ antennas are uniformly partitioned into $N_{\text{RF}}$ disjoint sub-arrays, such that each RF chain is exclusively connected to a subset of $N_{\textnormal{l}}/N_{\text{RF}}$ antennas. For simplicity, we assume that $N_{\textnormal{l}}/N_{\text{RF}}$ is an integer. The transmit antennas at each AP are configured as a uniform linear array. To mitigate mutual coupling while maintaining high spatial resolution, the inter-element spacing is set to $d = \lambda/2$, where $\lambda$ denotes the carrier wavelength. Consequently, the aggregate number of transmit antennas across the entire CF network is $N_\text{t} = L N_{\textnormal{l}}$. The $k$-th user is equipped with $N_k$ receive antennas, resulting in a total of ${N}_{\text{r}} = \sum_{k=1}^K N_k$ receive antennas for all users.  Each AP reuses its downlink communication waveform for sensing, and all APs receive the echo signal. The received sensing signal at the $l$-th AP is modeled as the superposition of mono-static and bi-static echo generated by the signals transmitted from all APs and reflected by the target. Let the sensing path is from the $l$-th transmitting AP to the $\bar{l}$-th receiving AP.

Each AP first estimates the imperfect CSI and forwards it via the fronthaul links to the CPU for centralized processing. Upon receiving the CSI feedback, the CPU performs joint precoding design to manage multi-user interference and optimize sensing performance. Subsequently, the processed signals are transmitted synchronously by all APs. In this ISAC framework, the communication signals also serve as probing signals for sensing, and each AP captures the echo signals reflected by the target and forwards them back to the CPU for joint sensing information extraction.

The imperfect CSI is modeled as $\mathbf{H} = \tilde{\mathbf{H}} + \bar{\mathbf{H}}$, where $\tilde{\mathbf{H}}$ and $\bar{\mathbf{H}}$ denote the estimated channel and estimation error matrices, respectively. The aggregate channel matrix is defined as $\mathbf{H} \triangleq [\mathbf{H}_1^{\text{T}}, \mathbf{H}_2^{\text{T}}, \dots, \mathbf{H}_K^{\text{T}}]^{\text{T}} \in \mathbb{C}^{{N}_{\text{r}} \times N_\text{t}}$, where $\mathbf{H}_k \in \mathbb{C}^{N_k \times N_\text{t}}$ represents the channel from all APs to user $k$. Specifically, $\mathbf{H}_k$ is partitioned as $\mathbf{H}_k \triangleq [(\mathbf{H}_k^1)^{\text{T}}, (\mathbf{H}_k^2)^{\text{T}}, \dots, (\mathbf{H}_k^L)^{\text{T}}]^{\text{T}}$, where $\mathbf{H}_k^l \in \mathbb{C}^{N_k \times N_{\textnormal{l}}}$ denotes the sub-channel matrix between AP $l$ and user $k$.



\begin{figure*}[!b]
	\centering
	\vspace*{8pt}
    	\hrulefill
	\vspace*{8pt}
	\begin{align}
  & R_{\textnormal{c},k} \triangleq \textnormal{log\ det}(1+\frac{|\sum_{l=1}^L\textbf H_k^l\textbf F_{\textnormal{RF}}^l\textbf F_{\textnormal{BB,c}}^l|^2}{ \quad  \sigma_k^2 +|\sum_{j\neq k}^{K}\sum_{l=1}^{L} \textbf H_k^l\textbf F_{\textnormal{RF}}^l\textbf F_{\textnormal{BB,p}j}^l+\sum_{l=1}^{L} \textbf H_k^l\textbf F_{\textnormal{RF}}^l\textbf F_{\textnormal{BB,p}k}^l |^2}).\label{eq:4}\tag{4}\\
	&R_{\textnormal{p},k} \triangleq \textnormal{log\ det}(1+\frac{|\sum_{l=1}^L\textbf H_k^l\textbf F_{\textnormal{RF}}^l\textbf {F}_{\textnormal{BB,p}k}^l|^2}{ \quad  \sigma_k^2 +|\sum_{j\neq k}^{K}\sum_{l=1}^{L} \textbf H_k^l\textbf F_{\textnormal{RF}}^l\textbf F_{\textnormal{BB,p}j}^l|^2}).\label{eq:5}\tag{5}
	\end{align}

\end{figure*}
\subsection{Communication Model}
The system adopts the RS scheme, where the message intended for each user is partitioned into common parts and private parts. The common parts of all $K$ users are combined into a single common stream ${s}_{\text{c}}$, while the private signal vector is $\mathbf{s}_{\text{p}} = [\mathbf{s}_1^{\text{T}}, \mathbf{s}_2^{\text{T}}, \dots, \mathbf{s}_K^{\text{T}}]^{\text{T}} \in \mathbb{C}^{{N}_{\text{r}} \times 1}$. The aggregate signal vector is denoted by $\mathbf{s} = [{s}_{\text{c}}, \mathbf{s}_{\text{p}}^{\text{T}}]^{\text{T}} \in \mathbb{C}^{({N}_{\text{r}} + 1) \times 1}$.

The RS precoding matrix is defined as $\mathbf{F} = [\mathbf{F}_{\text{c}}, \mathbf{F}_{\text{p}}] \in \mathbb{C}^{N_\text{t} \times ({N}_{\text{r}} + 1)}$. The common precoding vector is given by $\mathbf{F}_{\text{c}} = \mathbf{F}_{\text{RF}} \mathbf{F}_{\text{BB,c}} \in \mathbb{C}^{N_\text{t} \times 1}$, where $\mathbf{F}_{\text{RF}} \in \mathbb{C}^{N_\text{t} \times LN_{\text{RF}}}$ and $\mathbf{F}_{\text{BB,c}} \in \mathbb{C}^{LN_{\text{RF}} \times 1}$ denote the analog precoding and the common digital precoding, respectively. Similarly, the private precoding matrix is formulated as $\mathbf{F}_{\text{p}} = \mathbf{F}_{\text{RF}} \mathbf{F}_{\text{BB,p}} \in \mathbb{C}^{N_\text{t} \times {N}_{\text{r}}}$, where $\mathbf{F}_{\text{BB,p}} \in \mathbb{C}^{LN_{\text{RF}} \times {N}_{\text{r}}}$ is the private digital precoding matrix. The transmitted symbol vector $\textbf x \in \mathbb{C}^{N_\text{t} \times 1}$ is then expressed as
\begin{equation}
\begin{aligned}
\textbf x =\sum_{l=1}^L (\sum_{k=1}^{K} \textbf F_{\textnormal{RF}}^l \textbf {F}_{\textnormal{BB,p}k}^l\textbf s_k+ \textbf F_{\textnormal{RF}}^l\textbf F_{\textnormal{BB,c}}^l s_{\textnormal{c}})
\label{eq:1},
\end{aligned}
\end{equation}
where $\textbf F_{\textnormal{RF}}^l\in{\mathbb{C}^{N_{{l}}\times{N_{\text{RF}}}}}$ denotes the analog precoding at AP $l$, $\textbf F_{\textnormal{RF}} = [{(\textbf F_{\textnormal{RF}}^1)}^{\text{T}}, {(\textbf F_{\textnormal{RF}}^2)}^{\text{T}},  {(\textbf F_{\textnormal{RF}}^L)}^{\text{T}}]^{\text{T}}$, $\textbf F_{\textnormal{BB,p}k}^l\in{\mathbb{C}^{N_{\text{RF}}\times{N_k}}}$ denotes private signal precoding for user~$k$ at AP $l$, $\textbf F_{\textnormal{BB,p}} = [\textbf F_{\textnormal{BB,p}1}, \textbf F_{\textnormal{BB,p}2},\dots, \textbf F_{\textnormal{BB,p}K}]$, $\textbf F_{\textnormal{BB,p}k} = [{(\textbf F_{\textnormal{BB,p}k}^1)}^{\text{T}},{( \textbf F_{\textnormal{BB,p}k}^2)}^{\text{T}},\dots, {(\textbf F_{\textnormal{BB,p}k}^L)}^{\text{T}}]^{\text{T}}\in{\mathbb{C}^{LN_{\text{RF}}\times{N_k}}}$.

The downlink received signal is given by Eq.~\eqref{eq:2}  at the top of this page. Let $\mathbf{y}_k \in \mathbb{C}^{N_k \times 1}$ be the received signal of user $k$. The vector $\mathbf{n}_k \in \mathbb{C}^{N_k \times 1}$ denotes the additive white Gaussian noise (AWGN). In the RS architecture, $\mathbf{S}_{\text{c},k}$ and $\mathbf{S}_{\text{p},k}$ represent the common and private information components intended for user $k$ through the wireless channel, respectively. To characterize the interference environment, $\mathbf{I}_{\text{c},k}$ denotes the interference of common streams and $\mathbf{I}_{\text{p},k}$ denotes the interference  affecting the private signals due to multi-user crosstalk and inter-layer interference. By concatenating the observations from all $K$ users, the aggregate received signal and the composite noise vectors are defined as $\mathbf{y} \triangleq [\mathbf{y}_1^{\text{T}}, \dots, \mathbf{y}_K^{\text{T}}]^{\text{T}} \in \mathbb{C}^{N_{\textnormal{r}} \times 1}$ and $\mathbf{n} \triangleq [\mathbf{n}_1^{\text{T}}, \dots, \mathbf{n}_K^{\text{T}}]^{\text{T}} \in \mathbb{C}^{N_{\textnormal{r}} \times 1}$, respectively. For all users, the received signal model is formulated as
\begin{equation}
\begin{aligned}
     \textbf y &= \textbf H\textbf F_{\textnormal{RF}}\textbf F_{\text{BB,p}}\textbf s+\textbf H\textbf F_{\textnormal{RF}}\textbf F_{\textnormal{BB,c}}\textbf s_{\textnormal{c}}+\textbf n ,
\end{aligned}
\label{eq:3}
\tag{3}
\end{equation}
where the data streams are assumed to be independent of each other, so $\mathbb{E}[\textbf s \textbf s^{\text{H}}] = \textbf I$.

At the receiver, successive interference cancellation is employed to decode and subsequently subtract the common stream from the composite received signal $\mathbf{y}_k$, which contains both common and private components. This procedure effectively mitigates common stream interference, facilitating the detection of the private signals $\mathbf{s}_{\text{p},k}$ in the second stage of the decoding process. The RS MIMO systems contain the common stream rate $R_{\text{c},k}$ as shown in Eq.~\eqref{eq:4} and the private stream rate $R_{\text{p},k} $ as shown in Eq.~\eqref{eq:5}. Eq.~\eqref{eq:4} and Eq.~\eqref{eq:5} are provided at the bottom of this page.

In this paper, we consider the scenario with imperfect CSI. Therefore, the ESR is adopted as the performance metric to evaluate the system. Given a channel estimate $\tilde{\mathbf{H}}$, the ESR of the system is formulated as follows 
\begin{equation}
\begin{aligned}
    \text{S}_{\textnormal{r}} = \min\limits_{k \in [1,K]}\mathbb{E}_{\tilde{\textbf H}}{[\text{R}_{\textnormal{c},k}]}+\mathbb{E}_{\tilde{\textbf H}}{[\text{R}_{\textnormal{p}}]},
\end{aligned}
\label{eq:6}
\tag{6}
\end{equation}
where $R_{\text{p}}=\sum_{k=1}^KR_{\text{p},k}$, denotes the private precoding rate for all users.
\subsection{Sensing Model}
In our case, the spatial coordinates of the $p$-th target are represented by the vector $\mathbf{L}_{\text{o}(p)} = [L_{\text{ox}(p)}, L_{\text{oy}(p)}]^{\text{T}} \in \mathbb{R}^{2}$, $p \in \mathcal{P}$, $ \mathcal{P}\triangleq\{1, 2, \dots,P\}$. Consequently, the joint location vector for the entire set of $P$ targets is given by

\begin{equation}
\begin{aligned}
\mathbf{L}_{\text{o}} = [(\mathbf{L}_{\text{o}(1)})^{\text{T}}, (\mathbf{L}_{\text{o}(2)})^{\text{T}}, \dots, (\mathbf{L}_{\text{o}(P)})^{\text{T}}]^{\text{T}} \in \mathbb{R}^{2P}
\end{aligned},
\label{eq:7}
\tag{7}
\end{equation}
which are related to the delay parameters $\tau_{p}^{l\bar{l}}$, the path-loss coefficients $\beta_{p}^{l\bar{l}}$, and the radar cross section coefficients $\gamma_p^{l,\bar{l}}$ for the sensing path from AP $l$ to AP $\bar{l}$ via target $p$,   where $l, \bar{l} \in \{1, \dots, L\}$ denote the indices of the transmitting and receiving APs, respectively. To maintain analytical tractability and simplify the subsequent precoding design, we initially construct the Fisher information matrix (FIM) for the joint parameter set comprising $\{\tau_p^{l,\bar{l}}\}$, $\{\beta_p^{l,\bar{l}}\}$, and $\{\gamma_p^{l,\bar{l}}\}$. 
We define the parameter vectors as follow
\begin{equation}
\begin{aligned}
   \boldsymbol{\eta} &= \left[{\textbf{L}}_o^{\text{T}}, \left(\boldsymbol{\gamma}^{\text{R}}\right)^{\text{T}}, \left(\boldsymbol{\gamma}^{\text{I}}\right)^{\text{T}}\right]^{\text{T}},\quad
   \boldsymbol{\theta} = \left[\boldsymbol{\tau}^{\text{T}}, \boldsymbol{\beta}^{\text{T}}, \left(\boldsymbol{\gamma}^{\text{R}}\right)^{\text{T}}, \left(\boldsymbol{\gamma}^{\text{I}}\right)^{\text{T}}\right]^{\text{T}}
\end{aligned},
\tag{8}
\end{equation}
where $\boldsymbol{\gamma}^{\text{R}} = [({\gamma_1^{1,1}})^{\text{R}}, ({\gamma_1^{1,2}})^{\text{R}},   \cdots, ({\gamma_p^{L,L}})^{\text{R}}, \cdots,({\gamma_P^{L,L}})^{\text{R}}]^{\text{T}}$ and  $\boldsymbol{\gamma}^{\text{I}} = [({\gamma_1^{1,1}})^{\text{I}}, ({\gamma_1^{1,2}})^{\text{I}},   \cdots, ({\gamma_p^{L,L}})^{\text{I}}, \cdots,({\gamma_P^{L,L}})^{\text{I}}]^{\text{T}}$ are the real and imaginary parts of the radar cross section coefficients corresponding to the $L^2$ propagation paths among the $L$ APs, respectively, and $\boldsymbol{\tau} = [\tau_1^{1,1}, \tau_1^{1,2}, \cdots, \tau_p^{L,L}, \cdots, \tau_P^{L,L}]^{\text{T}}$ and $\boldsymbol{\beta} = [\beta_1^{1,1}, \beta_1^{1,2}, \cdots, \beta_p^{L,L},\cdots, \beta_P^{L,L}]^{\text{T}}$. The received sensing signal at the AP $l$ is modeled as

\begin{equation}
\begin{aligned}
&\mathbf{Y}_p^{l,\bar{l}}(t) = \,  \sqrt{\beta_{p}^{l\bar{l}}}\gamma_p^{l,\bar{l}} \mathbf{a}_{\text{r},p}^{\bar{l}}(\psi_{\text{r},p}^{\bar{l}}) {(\mathbf{a}_{\text{t},p}^l)}^{\text{T}}(\psi_{\text{t},p}^{l}) \textbf{x}_l (\bar{t} - \tau_p^{l,\bar{l}}) \\
& + \sum_{t \neq p} \sqrt{\beta_{t}^{l\bar{l}}}\gamma_t^{l,\bar{l}} \mathbf{a}_{\text{r}, t}^{\bar{l}}(\psi_{\text{r}, t}^{\bar{l}}) {\mathbf{a}_{\text{t},p}^l}^{\text{T}}(\psi_{\text{t}, t}^{l}) \textbf{x}_l (\bar{t} - \tau_p^{l,\bar{l}}) + \mathbf{N}^{l,\bar{l}}(\bar{t})
\end{aligned},
\label{eq:9}
\tag{9}
\end{equation} 
where $l \neq \bar{l}$. $\mathbf{a}_{\text{t},p}^l(\psi_{\text{t},p}^{l}) \in \mathbb{C}^{N_{\text{l}} \times 1}$ is the transmit steering vector of the $l$-th AP towards target $p$. $\mathbf{a}_{\text{r},p}^{\bar{l}}(\psi_{\text{r},p}^{\bar{l}}) \in \mathbb{C}^{N_{\text{r}} \times 1}$ is the receive steering vector of the $\bar{l}$-th AP from target $p$.
The sensing performance is evaluated based on the PEB in \cite{11165351}. The time delay $\tau_p^{l,\bar{l}}$, which
corresponds to the signal propagation from the $l$-th AP to the
target and then to the $\bar{l}$-th AP, can be written as

\begin{equation}
\begin{aligned}
 \tau_p^{l,\bar{l}} \nonumber &= \frac{\sqrt{ ({\rho}_l - \rho)^2n_x+({\rho}_{\bar{l}} - \rho)^2n_y}}{c} \\
&+\frac{\sqrt{(\rho_l - \rho)^2n_x + (\rho_{\bar{l}} - \rho)^2n_y}}{c} \nonumber \\
&\triangleq \frac{d_l + d_{\bar{l}}}{c}
\end{aligned},
\tag{10}
\end{equation}
where $\rho_l$ and $\rho$ denote the coordinates of the $l$-th AP and the target, respectively, and $(\rho_l - \rho)^2$ represents the squared Euclidean distance between them. $n_x= [1,0]^{\text{T}}, n_y= [0,1]^{\text{T}}$. $d_l$ and $d_{\bar{l}}$  denote the distance from the target $p$ to the $l$-th and $\bar{l}$ AP. 



According to \cite{11091535}, the FIM  is shown as
\begin{equation}
  \begin{aligned}
\mathbf{J}(\boldsymbol{\theta}) = \frac{2}{\sigma_n^2}
\begin{bmatrix}
\mathbf{S} & \mathbf{0} & \mathbf{0} & \mathbf{0} \\
\mathbf{0} & \mathbf{V} & \mathbf{E}^{\text{R}} & \mathbf{E}^{\text{I}} \\
\mathbf{0} & (\mathbf{E}^{\text{R}})^{\text{T}} & \boldsymbol{\Lambda} & \mathbf{0} \\
\mathbf{0} & (\mathbf{E}^{\text{I}})^{\text{T}} & \mathbf{0} & \boldsymbol{\Lambda}
\end{bmatrix}_{(4L^2)\times(4L^2)}
 \end{aligned}  ,
 \tag{11}
\end{equation}
where $\mathbf{S}$, $\mathbf{V}$, $\mathbf{E}^{\text{R}}$, $ \mathbf{E}^{\text{I}}$ and $\boldsymbol{\Lambda}$ are all $L^2 \times L^2$ diagonal
matrices, respectively. 
\begin{equation}
  \begin{aligned}
&\mathbf{S}_{(\bar{l}-1)L+l,(\bar{l}-1)L+l} = 4\pi^2 \beta_{p}^{l\bar{l}}|\gamma_p^{l,\bar{l}}|^2 \sum_{k}  |{\mathbf{a}_{\text{t},p}^l}^{\text{T}}(\psi_{\text{t},p}^{l})\mathbf{F}_{\text{RF}}^l\mathbf{F}_{\text{BB},k}^l|^2 \\&
\mathbf{V}_{(\bar{l}-1)L+l,(\bar{l}-1)L+l} = \frac{1}{4\beta_p^{l,\bar{l}}} |\gamma_p^{l,\bar{l}}|^2 \sum_{k} |{\mathbf{a}_{\text{t},p}^l}^{\text{T}}(\psi_{\text{t},p}^{l})\mathbf{F}_{\text{RF}}^l\mathbf{F}_{\text{BB},k}^l|^2 \\&
\mathbf{E}^{\text{R}}_{(\bar{l}-1)L+l,(\bar{l}-1)L+l} = \frac{1}{2} {(\gamma_p^{l,\bar{l}}})^{\text{R}} \sum_{k} |{\mathbf{a}_{\text{t},p}^l}^{\text{T}} (\psi_{\text{t},p}^{l})\mathbf{F}_{\text{RF}}^l\mathbf{F}_{\text{BB},k}^l|^2\\&
\mathbf{E}^{\text{I}}_{(\bar{l}-1)L+l,(\bar{l}-1)L+l} = \frac{1}{2} {(\gamma_p^{l,\bar{l}})}^{\text{I}} \sum_{k} |{\mathbf{a}_{\text{t},p}^l}^{\text{T}}(\psi_{\text{t},p}^{l})\mathbf{F}_{\text{RF}}^l\mathbf{F}_{\text{BB},k}^l|^2 \\ &
\boldsymbol{\Lambda}_{(\bar{l}-1)L+l,(\bar{l}-1)L+l}= \beta_{p}^{l\bar{l}}\sum_{k} |{\mathbf{a}_{\text{t},p}^l}^{\text{T}}(\psi_{\text{t},p}^{l})\mathbf{F}_{\text{RF}}^l\mathbf{F}_{\text{BB},k}^l|^2 
 \end{aligned}.
 \tag{12}
\end{equation}


The FIM of estimating $\boldsymbol{\eta}$ can be expressed as

$$
\mathbf{J}(\boldsymbol{\eta}|\textbf{x}) = \mathbf{T}\mathbf{J}(\boldsymbol{\theta})\mathbf{T}^{\text{T}},
$$
where $\mathbf{T} = \frac{\partial \boldsymbol{\theta}}{\partial \boldsymbol{\eta}}$ is the Jacobian matrix, defined as
\begin{equation}
\begin{aligned}
\mathbf{T}& =
\begin{bmatrix}\frac{\partial}{\partial {L}_{o(x)}}\boldsymbol{\tau}^{\text{T}} & \frac{\partial}{\partial {L}_{o(x)}}(\boldsymbol{\beta})^{\text{T}} & \frac{\partial}{\partial {L}_{o(x)}}(\boldsymbol{\gamma}^{\text{R}})^{\text{T}} & \frac{\partial}{\partial {L}_{o(x)}}(\boldsymbol{\gamma}^{\text{I}})^{\text{T}} \\
\frac{\partial}{\partial {L}_{o(y)}}\boldsymbol{\tau}^{\text{T}} & \frac{\partial}{\partial {L}_{o(y)}}(\boldsymbol{\beta})^{\text{T}} & \frac{\partial}{\partial {L}_{o(y)}}(\boldsymbol{\gamma}^{\text{R}})^{\text{T}} & \frac{\partial}{\partial {L}_{o(y)}}(\boldsymbol{\gamma}^{\text{I}})^{\text{T}} \\
\frac{\partial}{\partial \boldsymbol{\gamma}^{\text{R}}}\boldsymbol{\tau}^{\text{T}} & \frac{\partial}{\partial \boldsymbol{\gamma}^{\text{R}}}(\boldsymbol{\beta})^{\text{T}} & \frac{\partial}{\partial \boldsymbol{\gamma}^{\text{R}}}(\boldsymbol{\gamma}^{\text{R}})^{\text{T}} & \frac{\partial}{\partial \boldsymbol{\gamma}^{\text{R}}}(\boldsymbol{\gamma}^{\text{I}})^{\text{T}} \\
\frac{\partial}{\partial \boldsymbol{\gamma}^{\text{I}}}\boldsymbol{\tau}^{\text{T}} & \frac{\partial}{\partial \boldsymbol{\gamma}^{\text{I}}}(\boldsymbol{\beta})^{\text{T}} & \frac{\partial}{\partial \boldsymbol{\gamma}^{\text{I}}}(\boldsymbol{\gamma}^{\text{R}})^{\text{T}} & \frac{\partial}{\partial \boldsymbol{\gamma}^{\text{I}}}(\boldsymbol{\gamma}^{\text{I}})^{\text{T}} 
\end{bmatrix}
\\
&= \begin{bmatrix}
\mathbf{X}_{2\times L^2} & \mathbf{Z}_{2\times L^2} & \mathbf{0}_{2\times 2L^2} \\
\mathbf{0}_{2L^2\times L^2} & \mathbf{0}_{2L^2\times L^2} & \mathbf{I}_{2L^2\times 2L^2}
\end{bmatrix}
\end{aligned},
\tag{13}
\end{equation}
and
\begin{equation}
\begin{aligned}
\mathbf{X} &= \frac{1}{c} \begin{bmatrix}
a_{1,1} & a_{1,2} & \cdots & a_{l,\bar{l}} & \cdots & a_{l,p}^L \\
b_{1,1} & b_{1,2} & \cdots & b_{l,\bar{l}} & \cdots & b_{L,L}
\end{bmatrix}\\
a_{l,\bar{l}} &= \frac{(\rho - \rho_l)n_x}{||\rho - \rho_l||} + \frac{(\rho - \rho_{\bar{l}})n_x}{||\rho - \rho_{\bar{l}}||} \\
b_{l,\bar{l}} &= \frac{(\rho - \rho_l)n_y}{||\rho - \rho_{\bar{l}}||} + \frac{(\rho - \rho_{\bar{l}})n_y}{||\rho - \rho_{\bar{l}}||}\\
\mathbf{Z} &= - \begin{bmatrix}
e_{1,1} & e_{1,2} & \cdots & e_{l,\bar{l}} & \cdots & e_{L,L} \\
f_{1,1} & f_{1,2} & \cdots & f_{l,\bar{l}} & \cdots & f_{L,L}
\end{bmatrix} \\
e_{l,\bar{l}} &= 2\beta_{l,\bar{l}} \left[ \frac{(\rho - \rho_{l})n_x}{(||\rho - \rho_{\bar{l}}||^2} + \frac{(\rho - \rho_{\bar{l}})n_x}{||\rho - \rho_{\bar{l}}||^2} \right]\\
f_{l,\bar{l}} &= 2\beta_{l,\bar{l}}  \left[ \frac{(\rho - \rho_{l})n_y}{||\rho - \rho_{\bar{l}}||^2} + \frac{(\rho - \rho_{\bar{l}})n_y}{||\rho - \rho_{\bar{l}}||^2} \right]
\end{aligned}.
\tag{14}
\end{equation}

\begin{equation}
\begin{aligned}
\text{CRB}_{(\textbf{L}_{\text{o}(p)})} =&
   \frac{c^2}{8\pi^2}\frac{1}{g_{xx}(\mathbf{F}_{\text{RF}}\mathbf{F}_{\text{BB}})g_{yy}(\mathbf{F}_{\text{RF}}\mathbf{F}_{\text{BB}}) - g_{xy}^2(\mathbf{F}_{\text{RF}}\mathbf{F}_{\text{BB}})}\\&
\begin{bmatrix}g_{yy}(\mathbf{F}_{\text{RF}}\mathbf{F}_{\text{BB}}) & -g_{xy}(\mathbf{F}_{\text{RF}}\mathbf{F}_{\text{BB}}) \\
-g_{xy}(\mathbf{F}_{\text{RF}}\mathbf{F}_{\text{BB}}) & g_{xx}(\mathbf{F}_{\text{RF}}\mathbf{F}_{\text{BB}})\end{bmatrix}
\end{aligned}.
\tag{15}
\end{equation}

\begin{equation}
\begin{aligned}
g_{xx}(\mathbf{F}_{\text{RF}}\mathbf{F}_{\text{BB}}) = &\sum_{\bar{l}}\sum_{l} a_{l,\bar{l}}^2 \beta_{p}^{l\bar{l}}|\gamma_p^{l,\bar{l}}|^2 \sum_{k} |\mathbf{a}_{t,l}^{\text{T}}(\psi_{\text{t},p}^{l})\mathbf{F}_{\text{RF}}^l\mathbf{F}_{\text{BB},k}^l|^2 \\
g_{yy}(\mathbf{F}_{\text{RF}}\mathbf{F}_{\text{BB}}) = &\sum_{\bar{l}}\sum_{l} b_{l,\bar{l}}^2 \beta_{p}^{l\bar{l}}|\gamma_p^{l,\bar{l}}|^2 \sum_{k} |\mathbf{a}_{t,l}^{\text{T}}(\psi_{\text{t},p}^{l})\mathbf{F}_{\text{RF}}^l\mathbf{F}_{\text{BB},k}^l|^2 \\
g_{xy}(\mathbf{F}_{\text{RF}}\mathbf{F}_{\text{BB}}) =& \sum_{\bar{l}}\sum_{l} a_{l,\bar{l}}b_{l,\bar{l}} \beta_{p}^{l\bar{l}}|\gamma_p^{l,\bar{l}}|^2\sum_{k} |\mathbf{a}_{t,l}^{\text{T}}(\psi_{\text{t},p}^{l})\mathbf{F}_{\text{RF}}^l\mathbf{F}_{\text{BB},k}^l|^2 
\end{aligned}.
\tag{16}
\end{equation}

Consequently, the PEB, which serves as a fundamental lower bound  is defined as

\begin{equation}
\begin{aligned}
\text{PEB}(\mathbf{L}_{\text{o}(p)}) = \sqrt{\text{Tr} \left( \text{CRB}_{(\textbf{L}_{\text{o}(p)})} \right)}
\end{aligned}.
\tag{17}
\end{equation}

\subsection{Problem Formulation}
In this paper, we aim to jointly design analog precoding $\textbf F_{\textnormal{RF}}$ and digital precoding $\textbf F_{\textnormal{BB}}$ to maximize the ESR $\text{S}_{\text{r}}$, and the problem is formulated as 

\begin{subequations}

\begin{align}
     \underset{\textbf{F}_{\textnormal{BB}}, \textbf{F}_{\textnormal{RF}}}{\max} &\quad \text{S}_{\text r}\label{eq:18} \tag{18} \\
    \text{s.t.}  \quad &\left\|\sum_{l=1}^L\textbf{F}_{\textnormal{RF}}^l\textbf{F}_{\textnormal{BB}}^l\right\|_{\text{F}}^2 = P  \label{eq:18a}\tag{18a}\\
    & \text{PEB}(\mathbf{L}_{\text{o}(p)}) \leq \text{PEB}_{\text{th}}\quad \forall p \in \mathcal{P} \label{eq:18b}   \tag{18b}   
 \\
    & |\mathbf{F}_{\mathrm{RF}}^l[i,j]|=1 \label{eq:18c}\tag{18c},
\end{align}
\end{subequations}
where \eqref{eq:18a} is the power budget constraint; \eqref{eq:18b} specifies the sensing performance threshold; \eqref{eq:18c} represents the analog precoding constraint.


\section{PROPOSED ALGORITHM}
\label{III}
This section proposes the algorithm to solve the optimization problem \eqref{eq:18}. It is generally difficult to simultaneously optimize the digital precoding $\mathbf{F}_{\mathrm{BB}}$ and the analog precoding $\mathbf{F}_{\mathrm{RF}}$, since they are coupled with each other in both the ESR objective function and sensing constraint. Moreover, the sensing constraint is non-convex, which significantly increases the difficulty of solving this problem. To address this challenge, we first derive a convex approximation for the sensing constraint. Next, we propose to optimize the digital precoding and analog precoding based on an AO framework. 

\subsection{ Convex Approximation for Sensing Constraint}
 According to \cite {11091535}, the optimization of \eqref{eq:18b} is non-convex, and the convex approximation is as follows.
\subsubsection*{Step 1: Equivalent Fractional Formulation}
The original non-convex sensing constraint \eqref{eq:18b} can be equivalently rewritten by utilizing the intermediate quadratic polynomial functions of $g_{xx}(\mathbf{F}_{\text{RF}}\mathbf{F}_{\text{BB}})$, $g_{yy}(\mathbf{F}_{\text{RF}}\mathbf{F}_{\text{BB}})$, and $g_{xy}(\mathbf{F}_{\text{RF}}\mathbf{F}_{\text{BB}})$ as
\begin{equation}
    \frac{g_{xx}(\mathbf{F}_{\text{RF}}\mathbf{F}_{\text{BB}}) + g_{yy}(\mathbf{F}_{\text{RF}}\mathbf{F}_{\text{BB}})}{g_{xx}(\mathbf{F}_{\text{RF}}\mathbf{F}_{\text{BB}})g_{yy}(\mathbf{F}_{\text{RF}}\mathbf{F}_{\text{BB}}) - g_{xy}^2(\mathbf{F}_{\text{RF}}\mathbf{F}_{\text{BB}})} \le \frac{8\pi^2}{c^2} \text{PEB}_{\text{th}}^2.
    \tag{19}
\end{equation}

\subsubsection*{Step 2: Transformation to Second-Order Cone Constraints}
To decouple the highly non-convex fractional constraint, we introduce auxiliary variables $\kappa_{xx}$, $\kappa_{yy}$, and $\kappa_{xy}$ to bound the quadratic terms $g_{xx}(\mathbf{F}_{\text{RF}}\mathbf{F}_{\text{BB}})$, $g_{yy}(\mathbf{F}_{\text{RF}}\mathbf{F}_{\text{BB}})$, and $g_{xy}(\mathbf{F}_{\text{RF}}\mathbf{F}_{\text{BB}})$, respectively. By further introducing $\tilde{t}_{xx} > 0$ and $\tilde{t}_{yy} > 0$, the constraint can be transformed into the following equivalent set of constraints.
\begin{equation}
\begin{aligned}
    & \tilde{t}_{xx}^{-1} + \tilde{t}_{yy}^{-1} \le \frac{8\pi^2}{c^2}& & \quad \text{PEB}_{\text{th}}^2 
     \tilde{t}_{xx}^{-1} \geq \frac{\kappa_{xx}}{\kappa_{xx} \kappa_{yy} - \kappa_{xy}^2}\\
    & \tilde{t}_{yy}^{-1} \geq \frac{\kappa_{yy}}{\kappa_{xx} \kappa_{yy} - \kappa_{xy}^2} & & \quad|g_{xy}(\mathbf{F}_{\text{RF}}\mathbf{F}_{\text{BB}})| \le \kappa_{xy} \\
    & 0 < \kappa_{xx} \le g_{xx}(\mathbf{F}_{\text{RF}}\mathbf{F}_{\text{BB}}) & & \quad0 < \kappa_{yy} \le g_{yy}(\mathbf{F}_{\text{RF}}\mathbf{F}_{\text{BB}})
\end{aligned}.
\tag{20}
\end{equation}

\subsubsection*{Step 3: Semidefinite Relaxation and Convexification}
During the AO process, the digital precoding $\mathbf{F}_{\text{BB}}$ is optimized for a given analog precoding $\mathbf{F}_{\text{RF}}$. We define a positive semidefinite digital covariance matrix $\mathbf{W} _k= \text{vec}(\mathbf{F}_{\text{BB},k})\text{vec}(\mathbf{F}_{\text{BB},k})^{\text{H}},\mathbf{F}_{\text{BB},k}\in \mathbb{C}^{LN_{\text{RF}}*(N_{\textnormal{r}}+1)}$. Let $\mathbf{A}_{xx}$, $\mathbf{A}_{yy}$, and $\mathbf{A}_{xy}$ be the constant block-diagonal coefficient matrices capturing the  $a_{l,\bar{l}}$, $\beta_p^{l,\bar{l}}$, $\xi_p^{l,\bar{l}}$, and  $\mathbf{a}_{t,l}(\psi_p^l)$. They are explicitly defined as
\begin{equation}
\begin{aligned}
    \mathbf{A}_{xx}^p &= \sum_{l} \text{blkdiag} \Big( a_{l,1}^2 \beta_p^{l,1} |\xi_p^{l,1}|^2 ({\mathbf{a}_{t,p}^1(\psi_p^1)})^*({\mathbf{a}_{t,p}^1}(\psi_p^1))^{\text{T}}\\
    &  \;\dots,a_{l,L}^2 \beta_{l,p}^L |\xi_{l,p}^L|^2 ({\mathbf{a}_{t,p}^L(\psi_p^L)})^*(\mathbf{a}_{t,p}^L(\psi_p^L))^{\text{T}} \Big) \\
    \mathbf{A}_{yy}^p &= \sum_{l} \text{blkdiag} \Big( \beta_p^{l,1} |\xi_p^{l,1}|^2 ({\mathbf{a}_{t,p}^1})^*(\psi_p^1)({\mathbf{a}_{t,p}^1})^{\text{T}}(\psi_p^1),  \\
    &   \;\dots, b_{l,L}^2 \beta_{l,p}^L |\xi_{l,p}^L|^2 ({\mathbf{a}_{t,p}^L})^*(\psi_p^L)({\mathbf{a}_{t,p}^L})^{\text{T}}(\psi_p^L) \Big) \\
    \mathbf{A}_{xy}^p &= \sum_{l} \text{blkdiag} \Big( a_p^{l,1} \beta_p^{l,1} |\xi_p^{l,1}|^2 ({\mathbf{a}_{t,p}^1})^*(\psi_p^1)({\mathbf{a}_{t,p}^1(\psi_p^1)})^{\text{T}}  \\
    &   \; \dots, a_{l,L}b_{l,L} \beta_{l,p}^L |\xi_{l,p}^L|^2 ({\mathbf{a}_{t,p}^L})^*(\psi_p^L)({\mathbf{a}_{t,p}^L})^{\text{T}}(\psi_p^L) \Big)
\end{aligned}.
\tag{21}
\end{equation}

The final reformulated constraints are given by
\begin{equation}
\begin{aligned}
    &\tilde{t}_{xx}^{-1} + \tilde{t}_{yy}^{-1} \le \frac{8\pi^2}{c^2} \text{PEB}_{\text{th}}^2\\
    & 0 < \kappa_{xx} \le \sum_k\text{Tr}(\mathbf{F}_{\text{RF}}^{\text{H}} \mathbf{A}_{xx}^p \mathbf{F}_{\text{RF}} \mathbf{W}_k) \\
    & 0 < \kappa_{yy} \le \sum_k\text{Tr}(\mathbf{F}_{\text{RF}}^{\text{H}} \mathbf{A}_{yy}^p \mathbf{F}_{\text{RF}} \mathbf{W}_k) \\
    & -\kappa_{xy} \le \sum_k\text{Tr}(\mathbf{F}_{\text{RF}}^{\text{H}} \mathbf{A}_{xy}^p \mathbf{F}_{\text{RF}} \mathbf{W}_k) \le \kappa_{xy} 
\end{aligned}.
\label{eq:22}
\tag{22}
\end{equation}

\subsection{Digital Precoding Design}
With the analog precoding fixed, we proceed to update the digital precoding. Subject to the constraints specific to the digital precoding, the subproblem is formulated as follows

\begin{align}
     \underset{\textbf{F}_{\textnormal{BB}}}{\max} &\quad \text{S}_{\text r} \tag{23}\label{eq:23}\\
    &\text{s.t.}  \quad (18a), (22)\nonumber.
\end{align}

Problem \eqref{eq:23} is non-convex since the ESR is not concave with respect to $\mathbf{F}_{\mathrm{BB}}$, owing to the coupled desired-signal and interference terms in the SINR expression. Therefore, directly solving problem \eqref{eq:23} is generally intractable. To overcome this hurdle, we transform \eqref{eq:23} into an equivalent optimization problem including slack variables as \cite{10902503}.
\begin{align}
 \max_{\mathcal{V}} \quad & \text{S}_{\text{r}} \label{eq:24}\tag{24}\\
\text{s.t.} \quad & (18a), (22) \nonumber \\
& {R_{\mathrm{c},k}} \le \frac{|\text{S}_{\text{c},k}|^2}{\varepsilon_{\mathrm{c},k}} \tag{24a} \label{eq:24a}\\
& R_{\mathrm{p},k} \le \frac{|\text{S}_{\text{p},k}|^2}{\varepsilon_{\mathrm{p},k}}\tag{24b} \label{eq:24b}\\
& \text{I}_{\text{c},k} \le {\varepsilon_{{\text{c}},k}} \tag{24c}\\
& \text{I}_{\text{p},k} \le \varepsilon_{\mathrm{p},k}\tag{24d}.
\end{align}
where incorporating $\textbf{F}_{\text{BB}}$ into the joint multi-variable set $\mathcal{V} = \{\textbf{F}_{\text{BB}}, \mathbf{R}, \boldsymbol{\epsilon}\}$ successfully decouples the implicit non-convex  $\text{S}_\text{r}$.  $\mathcal{V} = \{\mathbf{F}_{\mathrm{BB}}, \textbf{R},\boldsymbol{\varepsilon}\}$ is the set of variables with  $\boldsymbol{\varepsilon} = [\varepsilon_{\mathrm{p},1}, \cdots, \varepsilon_{\mathrm{c},K}], \textbf{R}=[R_{\text{p},1},\dots, R_{\text{c},K}]$. It should be emphasized that \(R_{c,k}\) and \(R_{p,k}\)
in problem \eqref{eq:24} are not treated as the original logarithmic
rate expressions.  The original non-convexity of the problem is equivalently transferred to the newly formulated boundary constraints \eqref{eq:24a} and \eqref{eq:24b}. We are motivated to adopt the successive convex approximation  approach. The surrogate optimization problem is subsequently formulated as

\begin{align}
 \max_{\mathcal{V}} \quad & \sum_{k=1}^K({R}_{\mathrm{c},k}+{R}_{\mathrm{p},k})\tag{25}\label{eq:25} \\
 \quad {s.t.}& (18a), (22) \nonumber \\
& {R}_{\mathrm{c},k} \le \text{g}_{\text{c},k}^{(i)}(\mathbf{F}_{\mathrm{BB,c}k}, \varepsilon_{\mathrm{c},k}) \tag{25a} \label{eq:25a}\\
&{R}_{\mathrm{p},k} \le \text{g}_{\text{p},k}^{(i)}(\mathbf{F}_{\mathrm{BB,p}k}, \varepsilon_{\mathrm{p},k})\tag{25b}\label{eq:25b}.
\end{align}

Here, the function $\text{g}_k(\mathbf{F}_{\text{BB},k}, \varepsilon)$ is the first-order Taylor expansion of $|\text{S}_{k}|^2 / \varepsilon$ given as 
\begin{equation}
\begin{aligned}
&{\text{g}^{(i)}_{\text{c}k}(\mathbf{F}_{\text{BB},ck}, \varepsilon)} = \frac{2\operatorname{Re}{((\text{S}^{(i)}_{\text{c},k}})^\text{H}\text{S}_{\text{c},k})}{\varepsilon_{\text{c},k}^{i}} - \frac{|\text{S}^{(i)}_{\text{c},k}|^2}{(\varepsilon_{\text{c},k}^{i})^2}\varepsilon_{\text{c},k}\\&
\text{g}^{(i)}_{\text{p}k}(\mathbf{F}_{\text{BB,p}k}, \varepsilon) = \frac{2\operatorname{Re}{((\text{S}^{(i)}_{\text{p},k}}^{i})^\text{H}\text{S}_{\text{p},k})}{\varepsilon_{\text{p},k}^i} - \frac{|\text{S}^{(i)}_{\text{p},k}|^2}{(\varepsilon_{\text{p},k}^{i})^2}{\varepsilon_{\text{p},k}},
\end{aligned}
\tag{26}
\end{equation}
where $i$ is the iteration number. Problem \eqref{eq:25} is a standard convex optimization problem. By solving this surrogate optimization problem and updating the local points iteratively with the solution obtained from the previous iteration, the successive convex
approximation approach guarantees convergence to a local optimal solution to the original problem \eqref{eq:23}.  


\subsection{Analog Precoding Design}


The digital precoding is fixed, we proceed to update the analog precoding. Considering the constraints specific to the analog precoding, the subproblem can be formulated as follows

\begin{align}
    \max_{\mathbf{F}_{\text{RF}}} \quad & \text{S}_{\text{r}} - \eta \left( P - \left\| \sum_{l=1}^L \mathbf{F}_{\text{RF}}^l \mathbf{F}_{\text{BB}}^l \right\|_{\text{F}}^2 \right)^2\nonumber\label{eq:27}\tag{27} \\
    \text{s.t.}  \quad &(21c) ,\nonumber
\end{align}
where $\eta$ is the penalty factor for the power constraint deviation.

The Euclidean gradient of the sum rate is given by
\begin{equation}
    \nabla \mathrm{Sr} =  \sum_{k=1}^K \nabla R_{p,k} + \sum_{k=1}^K \nabla {R}_{c,k}, \tag{28}\label{eq:28}
\end{equation}

\begin{equation}
    \nabla R_{\text{p},k} = \frac{1}{\ln(2)} \frac{\textbf{B}_k \nabla \textbf{A}_k - \textbf{A}_k \nabla \textbf{B}_k}{\textbf{A}_k \textbf{B}_k} \tag{29},
\end{equation}
\begin{equation}
    \nabla {R}_{\text{c},k} = -\frac{1}{\ln(2)} \frac{ \sum_{k} \left( \left(\frac{\textbf{A}_k}{\textbf{C}_k}\right)^{\beta - 1} \frac{\textbf{C}_k \nabla \textbf{A}_k - \textbf{A}_k \nabla \textbf{C}_k}{\textbf{C}_k^2} \right)}{\sum_{k } \left(\frac{\textbf{A}_k}{\textbf{C}_k}\right)^\beta}\tag{30},
\end{equation}

and $\textbf{A}_k, \textbf{B}_k$, and $\textbf{C}_k$ are defined as
\begin{align}
    \textbf{A}_k &= \mathbf{\overline{\tilde{H}}}_k^{\text{H}} \mathbf{F}_{\text{RF}} \mathbf{F}_{\text{c},k}^- \mathbf{F}_{\text{RF}}^{\text{H}} \mathbf{\overline{\tilde{H}}}_k, \tag{31} \\
    \textbf{B}_k &= \mathbf{\overline{\tilde{H}}}_k^{\text{H}} \mathbf{F}_{\text{RF}} \mathbf{F}_k^- \mathbf{F}_{\text{RF}}^{\text{H}} \mathbf{\overline{\tilde{H}}}, \tag{32} \\
    \textbf{C}_k &= \mathbf{\overline{\tilde{H}}}_k^{\text{H}} \mathbf{F}_{\text{RF}} \mathbf{F} \mathbf{F}_{\text{RF}}^{\text{H}} \mathbf{\overline{\tilde{H}}}_k, \tag{33}\label{eq:33}
\end{align}
where $\mathbf{F} = \mathbf{F}_{\text{BB}}\mathbf{F}_{\text{BB}}^{\text{H}}$, $\mathbf{F}_{\text{c},k}^- = \mathbf{F} - \mathbf{F}_{\text{c},k}\mathbf{F}_{\text{c},k}^{\text{H}}$, and $\mathbf{F}_k^- = \mathbf{F}_{\text{c},k}^- - \mathbf{F}_{\text{{p}}k}(\mathbf{F}_{\text{{p}}k})^{\text{H}}$. Consequently, the gradients of these variables are evaluated as
\begin{align}
    \nabla  \textbf{A}_k &= \mathbf{H}_k \mathbf{H}_k^{\text{H}} \mathbf{F}_{\text{RF}} \mathbf{F}_{\text{c},k}^- \tag{34}, \\
    \nabla  \textbf{B}_k &= \mathbf{H}_k \mathbf{H}_k^{\text{H}} \mathbf{F}_{\text{RF}} \mathbf{F}_k^- \tag{35} ,\\
    \nabla  \textbf{C}_k &= \mathbf{H}_k \mathbf{H}_k^{\text{H}} \mathbf{F}_{\text{RF}} \mathbf{F}. \tag{36}\label{eq:36}
\end{align}

The gradient of the penalty term is simply derived using $\nabla \left\|\sum_{l=1}^L\textbf{F}_{\textnormal{RF}}^l\textbf{F}_{\textnormal{BB}}^l\right\|_{\text{F}}^2 = 2\mathbf{F}_{\text{RF}}\mathbf{F}_{\text{BB}}$. It is worth noting that unlike the power constraint, which is handled via a penalty term in the objective function, the constant modulus constraint \eqref{eq:18c} does not need to be explicitly penalized. This is because the unit-modulus constraints inherently form a complex circle Riemannian manifold. The Riemannian conjugate gradient algorithm directly performs optimization over this manifold. The Riemannian conjugate gradient algorithm ensures that the updated analog precoding matrix strictly stays on the manifold surface at each iteration, naturally satisfying the constant modulus constraint without requiring additional penalty formulations. Utilizing this gradient information, the Riemannian conjugate gradient algorithm can be applied to find a local optimal solution to problem \eqref{eq:27}, thereby obtaining the analog precoding.

\begin{algorithm}[t]
    \caption{Hybrid Precoding based on CF-AO}
	\label{algorithm2} 
	\renewcommand{\algorithmicrequire}{\textbf{Input:}}
	\renewcommand{\algorithmicensure}{\textbf{Output:}}
	\begin{algorithmic}[1]
		\REQUIRE $\textbf s,\textbf H$ and $N_{\textnormal{r}}$;
		\ENSURE $\textbf F_{\textnormal{RF}} ,\textbf F_{\textnormal{BB}} $.

        \STATE Initialize $\textbf F_{\textnormal{RF}}^l$ satisfying the constraint \eqref{eq:18c} in APs;

        \STATE CPU receives $\overline{\tilde{\textbf H}}^l\textbf F_{\textnormal{RF}}^l$;

        \STATE The CPU unites the $\overline{\tilde{\textbf H}}^{l}\textbf F_{\textnormal{RF}}^l$ obtained by APs and computes $\overline{\tilde{\textbf H}}\textbf F_{\textnormal{RF}}$;

        \STATE Obtain $\textbf F_{\textnormal{BB,c}}$ by MMSE in CPU;

        \FOR{$\text{index}=1$ to $\text{Iter}$}

            \STATE Update $\textbf F_{\textnormal{BB,p}}$, constraints \eqref{eq:18a} and \eqref{eq:22};

        \ENDFOR 

        \STATE CPU transmits $\textbf F = [\textbf{F}_{\textnormal{BB,c}}^l\ \textbf{F}_{\textnormal{BB,p}}^l]$ to APs;

        \IF{\eqref{eq:18}}

            \RETURN $\textbf F_{\textnormal{RF}} ,\textbf F_{\textnormal{BB}}$

        \ELSE

            \STATE Repeat steps $2$--$9$.

        \ENDIF

        \RETURN $\textbf F_{\textnormal{RF}} ,\textbf F_{\textnormal{BB}}$
	\end{algorithmic} 
\end{algorithm}

\subsection{Fronthaul Overhead and Complexity Analysis}
For the aforementioned convex optimization procedure, the analog precoding design is executed at the APs, and the CSI is subsequently forwarded to the CPU. However, the prohibitive fronthaul transmission overhead remains a critical challenge in CF massive MU-MIMO systems. To mitigate this issue, we employ analog precoding to compress the channel matrices and adopt a partially-connected architecture to reduce the transmission overhead. Upon receiving the compressed channel information, the CPU performs the optimization for the digital precoding. Specifically, the compressed channel matrices, represented by $\overline{\tilde{\mathbf{H}}_l} \mathbf{F}_{\mathrm{RF}}^l$, are transmitted from the APs to the CPU. 

By leveraging channel matrix compression via analog precoding and incorporating a partially-connected architecture, both the fronthaul transmission overhead and the computational complexity are significantly mitigated. The detailed analysis of the fronthaul overhead and computational complexity is presented as follows.

\textit{Fronthaul transmission  overhead}: To resolve the severe fronthaul bottleneck caused by forwarding the full-dimensional CSI $\overline{\tilde{\mathbf{H}}} \in \mathbb{C}^{N_{\mathrm{r}} \times N_{\mathrm{l}}}$ to the CPU, we propose an integrated mechanism combining analog precoding-based spatial compression with a partially-connected architecture. By projecting the high-dimensional CSI onto the analog precoding $\mathbf{F}_{\mathrm{RF}}$, APs generate a compressed equivalent channel $\overline{\tilde{\mathbf{H}}}_l \mathbf{F}_{\mathrm{RF}}^l$. This strategy completely decouples the overhead from the transmit antenna count, drastically shrinking the CPU transmission overhead from $N_{\mathrm{r}} \times N_{\mathrm{l}}$ to $N_{\mathrm{r}} \times N_{\mathrm{RF}}$. Specifically, setting $N_{\mathrm{RF}} = \frac{1}{2}N_{\text{l}}$ and $N_{\mathrm{RF}} < \frac{1}{8}N_{\text{l}}$ reduces transmission overhead by 50\% and over 87.5\%, respectively. Since fewer RF chains inevitably degrade performance, $N_{\mathrm{RF}}$ must be configured to balance overhead and system performance.

\textit{Complexity analysis}:  For the complexity of the proposed algorithm, solving problem \eqref{eq:25} directly via interior-point methods is computationally prohibitive due to the exponential constraints \eqref{eq:25a} and \eqref{eq:25b}. By applying the successive convex approximation approach, these constraints are equivalently recast into second-order cone constraints. This transforms \eqref{eq:25} into a standard second-order cone programming problem. While the complexity of this subproblem would typically scale as $\mathcal{O}(I_{\text{SCA}}(N_t(N_r + 1))^{3.5})$ in a fully connected system, the proposed partially connected architecture reduces this to $\mathcal{O}(I_{\text{SCA}}(N_{\text{RF}}(N_r + 1))^{3.5})$, where $I_{\text{SCA}}$ is the number of iterations.

The computational complexity of solving problem \eqref{eq:27} via the Riemannian conjugate gradient algorithm is dominated by the gradient evaluation in Eq.~\eqref{eq:33}. Computing $\mathbf{C}_k$ involves matrix multiplications with complexities of $\mathcal{O}(N_t^2(N_r + 1))$ and $\mathcal{O}(N_t^2)$. However, by exploiting the reduced dimensionality of the partially connected transmission matrices, these terms are effectively reduced to $\mathcal{O}(N_{\text{RF}}^2(N_r + 1))$ and $\mathcal{O}(N_{\text{RF}}^2)$, respectively. Since $\mathbf{C}_k$ must be evaluated for all $K$ UEs, the overall complexity for the analog precoding update is reduced from $\mathcal{O}(I_{\text{RCG}}((N_r + 2)N_t^2)K)$ to $\mathcal{O}(I_{\text{RCG}}(N_r + 2)N_{\text{RF}}^2K)$, where $I_{\text{RCG}}$ denotes the number of iterations. The  $\textbf{F}_{\text{BB}}$  requires $\mathcal{O}(I_{\text{F}}(N_{\text{r}}^3 + N_{\text{RF}}N_{\text{r}}^2 + N_{\text{r}}N_{\text{RF}}^2))$ ), where  $I_{\text{F}}$ is the number of iteration. Consequently, the total computational complexity is $\mathcal{O}(I_{\text{SCA}}(N_{\text{RF}}(N_r + 1)^{3.5})+I_{\text{F}}(N_{\text{r}}^3 + N_{\text{RF}}N_{\text{r}}^2 + N_{\text{r}}N_{\text{RF}}^2))$.

\section{Proposed Low Complexity Digital Algorithm}
\label{IV}
In our system, deriving the optimal digital precoding $\text{F}_{\text{BB}}$ via the aforementioned AO framework inherently entails high computational complexity. Furthermore, practical communication environments are highly dynamic.  Inevitable fluctuations in the number of active users and continuous channel variations dictate that the entire optimization process must be executed repeatedly to update the high-dimensional hybrid precoding matrices. This frequent need for recalculation causes the accumulated computational complexity to become prohibitive for real-time implementation. Therefore, developing a low-complexity digital precoding algorithm tailored for problem \eqref{eq:18} whose computational complexity is fundamentally robust and insensitive to continuous channel updates and varying user scales, emerges as a crucial research focus for practical system deployment.
\subsection{MB RS Algorithm}

\begin{figure}[t]   
\centering
\includegraphics[width=0.48\textwidth]{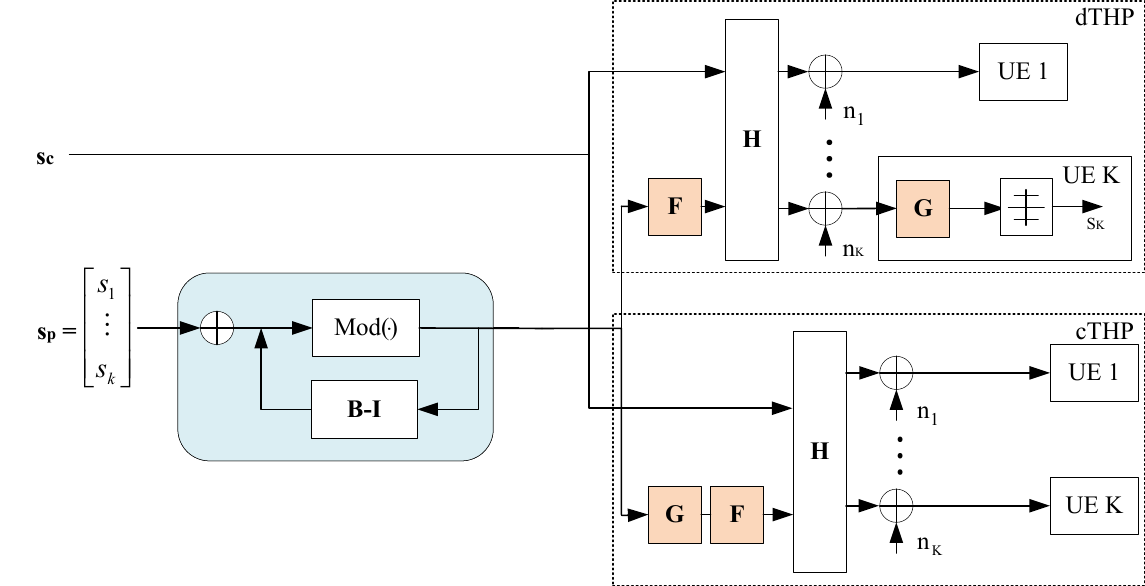}
\caption{Proposed RS-THP structures.$(a)$ Decentralized THP:  $\textbf G$ is separately placed at the receivers.
$(b)$ Centralized THP: $\textbf G$ is placed at the transmitter.}
\label{fig:3}
\end{figure}
To begin with, a channel preprocessing step is applied prior to tackling the optimization problem \eqref{eq:18}. Crucially, this operation remains entirely decoupled from and does not alter the subsequent optimization procedure. In CF massive MU-MIMO ISAC systems, THP demonstrates superior effectiveness in mitigating multi-user interference (MUI)\cite{b38}. The THP uses three filters, namely the feedforward filter $\overline{\tilde{\textbf F}}\in {\mathbb{C}^{{N_{\textnormal{t}}}\times N_{\textnormal{r}}}}$, which partially removes MUI, feedback filter $\overline{\tilde{\textbf B}}\in {\mathbb{C}^{{N_{\textnormal{r}}}\times N_{\textnormal{r}}}}$, which is a lower triangular matrix, and weighting matrix $\overline{\tilde{\textbf G}}\in {\mathbb{C}^{{N_{\textnormal{r}}}\times N_{\textnormal{r}}}}$, which contains weighting factors for each stream \cite{b23}.

We know that each AP transmits $\overline{\tilde{\textbf H}}^l$ to CPU, $\overline{\tilde{\textbf H}}^l = [(\overline{\tilde{\textbf H}}_k^1)^{\text{T}}, (\overline{\tilde{\textbf H_k}}^2)^{\text{T}}, ...,(\overline{\tilde{\textbf H}}_k^l)^{\text{T}}]^{\text{T}}$, the QR decomposition of  $\overline{\tilde{\textbf H}}_k$ shown below
\begin{equation}
\begin{aligned}
      \overline{\tilde{{\textbf H}}}_k^{\text{T}} &= \overline{\tilde{\textbf Q}}_k\overline{\tilde{\textbf R}}_k, \  \forall k,
\end{aligned}
\tag{37}
\end{equation}
where $\overline{\tilde{{\textbf H}}}_k$ denotes $\tilde{\textbf H}_k$ after undergoing the MB transformation. The order of the symbols affects the performance of the proposed RS MMSE-THP. Therefore, we further improve the performance by optimizing the ordering. The process of MB to $\overline{\tilde{{\textbf H}}}$ transformation is as follows

\begin{equation}
\begin{aligned}
     {\textbf M_{\text{u},1}}= \textbf I_k.\\
\end{aligned}
\tag{38}
\label{eq:38}
\end{equation}
\begin{equation}
\begin{aligned}
     {\textbf M_{\text{u},i}}= \begin{bmatrix}
\textbf I_{i-2} &0_{i-2,K-i+2}  \\
0_{K-i+2,i-2} &\prod_i^{\text{u}}
\end{bmatrix} ,2 < i\leq K\\
\end{aligned}.
\label{eq:39}
\tag{39}
\end{equation}

The matrix $\textbf M_{\text{u},i}$ represents the user $i$-th ordering mode, and the matrix $\prod_i^{\text{u}}\in {\mathbb{C}^{{(K-i+2)}\times {(K-i+2)}}}$ exchanges the user order. According to Eq.~\eqref{eq:39}, the equation can be rewritten as
\begin{equation}
\begin{aligned}
     {\textbf M_{\textbf s_k,1}}= \textbf I_{N_k}.\\
\end{aligned}
\tag{40}
\end{equation}
\begin{equation}
\begin{aligned}
     {\textbf M_{\textbf s_k,j}}= \begin{bmatrix}
\textbf I_{j-2} &0_{j-2,N_{\text{r}}-j+2}  \\
0_{N_{\text{r}}-j+2,j-2} &\prod_j^{\textbf s_k}\\
\end{bmatrix} ,2 < j\leq N_{\text{r}}\\
\end{aligned},
\tag{41}
\end{equation}
where $\textbf M_{\textbf s_k,j}$ denotes the ordering pattern of the $j$-th symbol of the $k$-th user, $\prod_j^{\textbf s_k}\in {\mathbb{C}^{{(N_{\text{r}}-j+2)}\times {(N_{\text{r}}-j+2)}}}$ exchanges the order of symbols of the $k$-th user. Since the users of our system have the same receiving antenna, we can use the product of $\textbf M_{\text{u},i}$ and $\textbf M_{\textbf{s}_k,j}$ together as the system sorting pattern
\begin{equation}
\begin{aligned}
     {\textbf M_{B}^{(i,j)}}= \textbf M_{\text{u},i}\otimes\textbf M_{\textbf s_k,j}, 1\leq B\leq B_{\text{max}}
\end{aligned},
\tag{42}
\end{equation}
$B_{\text{max}}$ denotes the maximum number of branches. MB cTHP and MB dTHP can achieve branch performance with $2$ or $4$ branches, ensuring lower complexity.

Considering the MB architecture, Eq.~\eqref{eq:6} can be rewritten as
\begin{equation}
\begin{aligned}
     {\textbf M_{\text{o}}}= \underset{B}{\max}(\underset{k\in[1,K]}{\min}\mathbb{E}[R_{\text{c},k}|\tilde{\textbf {H}}(\textbf{M}_B)]+\mathbb{E}[R_{\text{p}}|\tilde{\textbf {H}}(\textbf{M}_B)])
\end{aligned}.
\label{eq:43}
\tag{43}
\end{equation}

Once the optimal path ordering is made, the channel estimates are reordered, $\overline{\tilde{{\textbf H}}} = \textbf M_{\text{o}}\tilde{\textbf H}$. 

Within the RS precoding framework, the common and private digital precoding matrices are explicitly decoupled into adjustable power allocation variables.  Consequently, the constraints \eqref{eq:22} can be rigorously rewritten as
\begin{equation}
\begin{aligned}
   0   < \kappa_{xx} \le &\text{Tr}(\mathbf{F}_{\text{RF}}^{\text{H}} \mathbf{A}_{xx}^p \mathbf{F}_{\text{RF}} \mathbf{W}_{\text{c}}\text{P}_{\text{c}})\\&+\sum_k\text{Tr}(\mathbf{F}_{\text{RF}}^{\text{H}} \mathbf{A}_{xx}^p \mathbf{F}_{\text{RF}} \mathbf{W}_{\text{p},k}\beta^{\text{cTHP}}) \\
     0 < \kappa_{yy} \le &\text{Tr}(\mathbf{F}_{\text{RF}}^{\text{H}} \mathbf{A}_{yy}^p \mathbf{F}_{\text{RF}} \mathbf{W}_{\text{c}}\text{P}_{\text{c}})\\&+\sum_k\text{Tr}(\mathbf{F}_{\text{RF}}^{\text{H}} \mathbf{A}_{yy}^p \mathbf{F}_{\text{RF}} \mathbf{W}_{\text{p},k}\beta^{\text{cTHP}})  \\
     -\kappa_{xy} \le &\text{Tr}(\mathbf{F}_{\text{RF}}^{\text{H}} \mathbf{A}_{xy}^p \mathbf{F}_{\text{RF}} \mathbf{W}_{\text{c}}\\&+\sum_k\text{Tr}(\mathbf{F}_{\text{RF}}^{\text{H}} \mathbf{A}_{xy}^p \mathbf{F}_{\text{RF}} \mathbf{W}_{\text{p},k}\beta^{\text{cTHP}})  \le \kappa_{xy}, \\
\end{aligned}
\tag{44}
\label{eq:44}
\end{equation}
where $\mathbf{W}_{\text{c}}= \text{vec}(\mathbf{F}_{\text{BB,c}})\text{vec}(\mathbf{F}_{\text{BB,c}})^{\text{H}}, \mathbf{W}_{\text{p},k}= \text{vec}(\mathbf{F}_{\text{BB,p}k})\text{vec}(\mathbf{F}_{\text{BB,p}k})^{\text{H}}$, $\beta^{\text{cTHP}}$ denote the power scaling factors that satisfy the transmit power constraints. The power of the common stream is $||\text p_{\text{c}}||^2=\delta \text{E}_{\text{Tr}}$, $\delta$ denotes the percentage of $\text{E}_{\text{Tr}}$. The system is equivalent to the conventional spatial division multiple access when $\delta = 0$. In ZF-THP,  $\beta^{\text{cTHP}}$ and $\beta^{\text{dTHP}}$ are denoted as
\begin{equation}
\begin{aligned}
 \beta^{\text{cTHP}} = \sqrt{\frac{P-||\text p_{\text{c}}||^2}{\sum_{m=1}^{N_{\textnormal{t}}}r_{L,L}^2}},
\end{aligned}
\tag{45}
\end{equation}
where $r_{L,L}$ denotes the diagonal elements of $\textbf G$. In MMSE-THP, $\beta^{\text{cTHP}}$ is denoted as
\begin{equation}
\begin{aligned}
 \beta^{\text{cTHP}} = \sqrt{\frac{P-||\text p_{\text{c}}||^2}{\text{Tr}(\overline{\tilde{\textbf G}}\tilde{\textbf Q}\overline{\tilde{\textbf Q}}^{\text{H}}\overline{\tilde{\textbf G}}^{\text{H}})}}.
\end{aligned}
\tag{46}
\end{equation}

As illustrated in Fig.~\ref{fig:3}, the proposed RS-THP architecture transmits a common signal alongside THP-precoding private signals. To substantially alleviate the hardware and computational complexity at the receivers, we adopt the centralized THP (cTHP) configuration, which integrates the scaling matrix $\mathbf{G}$ into the transmitter. $\overline{\tilde{\textbf Q}}_k\in {\mathbb{C}^{{N_{\textnormal{t}}}\times N_k}}$ is the unitary matrix and $\overline{\tilde{\textbf R}}_k\in {\mathbb{C}^{{N_k}\times N_k}}$ is the upper triangular matrix, along with the weighting matrix $\overline{\tilde{\textbf G}}$. In the classical ZF-THP, $\overline{\tilde{\textbf F}}_k^l\in {\mathbb{C}^{{N_{\mathrm{l}}}\times N_k}}$  is set as $\overline{\tilde{\textbf Q}}_k^l$,  $(\overline{{\tilde{\textbf B}}_k^l})^{-1} = {\overline{\tilde{\textbf R}}_k^{-l\text{T}}}{\overline{\tilde{\textbf G}}_k^{-l}} $. In contrast to the LQ decomposition based correlation coefficient investigated in \cite{b30}, we derive the correlation coefficient under QR decomposition. The information stream is partitioned into common and private components, which are subsequently mapped to cTHP and dTHP structures, respectively. The resulting coefficients for these two configurations differ, thereby leading to distinct performance outcomes.

\begin{equation}
\begin{aligned}
   {\overline{\tilde{\textbf B}}_k^{l(\text{cTHP}){-1}}} = {{{(\overline{\tilde{\textbf R_k}}^l)}^{-\text{T}}}}{{(\overline{\tilde{\textbf G}}_k^l)}^{-1}}.
\end{aligned}
\label{eq:47}
\tag{47}
\end{equation}


\begin{equation}
\begin{aligned}
   \tilde{\textbf G}_k^l = 
\begin{bmatrix}
r_{1,1}^{-1}   \\
&r_{2,2}^{-1}  \\
&& \ddots \\
&&&r_{N_k,N_{k}}^{-1} \\
\end{bmatrix}.
\end{aligned}
\label{eq:48}
\tag{48}
\end{equation}

By substituting the RS expressions in Eq.~\eqref{eq:47} - Eq.~\eqref{eq:48} into Eq.~\eqref{eq:2}, we rewrite Eq.~\eqref{eq:2} as follows
 
\begin{equation}
\begin{aligned}
   & \textbf y_k^{\text{cTHP}}\\&={ {\textbf H}_k\textbf{F}_{\text{RF}}\textbf F_{\textnormal{BB,c}}s_{\textnormal{c}}}+\beta^{\text{cTHP}}( {\textbf H}_k\textbf{F}_{\text{RF}}\textbf F_{\textnormal{BB,p}}\textbf s_{\textnormal{p}}+\textbf n_k) \\ &={\sum_{l=1}^L {\textbf H}_k^l\textbf{F}_{\text{RF}}^l\textbf F_{\textnormal{BB,c}}^l s_{\textnormal{c}}}
    +\beta^{\text{cTHP}}(\sum_{l=1}^L\overline{\tilde{\textbf G}}_k^l {\textbf H}_k^l\textbf F_{\textnormal{RF}}^l\textbf {F}_{\textnormal{BB,p}k}^l({\overline{\tilde{\textbf B}}_k^l})^{-1}\textbf s_k^l\\
    &+\sum_{j\neq k}^{K}\sum_{l=1}^{L} \overline{\tilde{\textbf G}}_j^l {\textbf H}_k^l\textbf F_{\textnormal{RF}}^l\textbf F_{\textnormal{BB,p}j}^l({\overline{\tilde{\textbf  B}}_k^l})^{-1}\textbf 
 s_j^l+\sum_{l=1}^{L}\overline{\tilde{\textbf G}}_k^l\textbf n_k)
\end{aligned}
\tag{49}.
\end{equation}


Through the $\overline{\tilde{\textbf F}}$, $\overline{\tilde{\textbf G}}$ and $\overline{\tilde{\textbf{B}}}$ filters in the THP,  Eq.~\eqref{eq:2} can be rewritten  as
\begin{equation}
\begin{aligned}
     \textbf y_k &= \textbf s_k+\sum_{l=1}^{L}\overline{\tilde{\textbf G}}_k^l\textbf n_k. \\
\end{aligned}
\tag{50}
\label{eq:50}
\end{equation}

To demonstrate the general applicability of our proposed MB RS framework, we first introduce the ZF-THP design as a low-complexity analytical baseline. While the ZF-THP scheme benefits from inherently lower computational complexity, it is evident from Eq.~\eqref{eq:50} that it retains residual interference due to its susceptibility to noise enhancement. To overcome this limitation and fully exploit the system potential, we subsequently extend the framework to the MMSE-THP scheme. The corresponding QR decomposition for MMSE-THP is formulated as
\begin{equation}
\begin{aligned} 
    \Check{ \textbf H}_t =[\overline{\tilde{{\textbf H}}}_{t+1} \ \gamma] = \Check{\textbf{Q}} _{t+1}\Check{\textbf{R}}_{t+1},\\
\end{aligned}
\tag{51}
\end{equation}
where $\gamma=\sigma_n^2 / \sigma_x^2$. 

             
            


\subsection{RS Update Algorithm}
In addition to alleviating the computational complexity through the partially-connected analog architecture, we further reduce the overall complexity for the digital precoding. We propose a framework where MMSE is employed for common precoding, while MMSE-THP is utilized for private precoding. This strategy, which distinguishes our approach from the method in \cite{b31}, enables the pre-calculated MMSE to be seamlessly reused in the subsequent derivation of private precoding. Consequently, the overall computational complexity is substantially reduced relative to the benchmark in \cite{b31}. Furthermore, we develop a low-complexity update algorithm for both common and private precoding, enabling the system to efficiently adapt to fluctuating user numbers without compromising performance.

At time $t+1$, the scenario of a user arriving can be represented as $\overline{\tilde{\textbf{H}}}_{t+1}=[\overline{\tilde{\textbf{H}}}_t(1:\tau,:), \overline{\tilde{\textbf{H}}}_{\tau}, \overline{\tilde{\textbf{H}}}_t(\tau+1:N_{\textnormal{r}},:)]$, $\textbf{H}_{t+1}$ can be rewritten as
\begin{equation}
\begin{aligned}
     &{\overline{\tilde{\textbf H}}_{t+1}}= [\overline{\tilde{\textbf H}}_t \ \overline{\tilde{\textbf{H}}}_{\tau}]\textbf{E}=\overline{\tilde{{\textbf H}}}_{t}^{\text{T}}(\overline{\tilde{{\textbf H}}}_{t} \ \overline{\tilde{{\textbf H}}}_{t}^{\text{T}}+ \gamma\textbf I)^{-1}\\
    &\overline{\tilde{{\textbf H}}}_{t} \ \overline{\tilde{{\textbf H}}}_{t}^{\text{T}}+ \gamma\textbf I=  \textbf{W}_t
\end{aligned}
\label{eq:52}
\tag{52}.
\end{equation}

\begin{equation}
\begin{aligned}
     {\textbf W_{t+1}}&= [\overline{\tilde{\textbf H}}_t \ \overline{\tilde{\textbf{H}}}_{\tau}][\overline{\tilde{\textbf H}}_t \ \overline{\tilde{\textbf{H}}}_{\tau}]^{\text{T}}+ \gamma\textbf I\\
     &=\overline{\tilde{\textbf H}}_t\overline{\tilde{\textbf H}}_t^{\text{T}}+\gamma\textbf I+\overline{\tilde{\textbf{H}}}_{\tau}\overline{\tilde{\textbf{H}}}_{\tau}^{\text{T}}=\textbf{W}_t+\overline{\tilde{\textbf{H}}}_{\tau}\overline{\tilde{\textbf{H}}}_{\tau}^{\text{T}}
\end{aligned}
\label{eq:53}
\tag{53}.
\end{equation}

To reduce the computational complexity of the update algorithm, the Sherman-Morrison formula is employed to update the MMSE.
\begin{equation}
\begin{aligned}
     {\textbf F_{\text{BB,c}(t+1)}}&= [ \frac{\textbf W_t}{1+\overline{\tilde{{\textbf H}}}_{\tau}^{\text{T}}\textbf{W}_t^{-1}\overline{\tilde{{\textbf H}}}_{\tau}}  \  \frac{\overline{\tilde{\textbf H}}_{\tau}\textbf{W}_t^{-1}}{1+\overline{\tilde{{\textbf H}}}_{\tau}^{\text{T}}\textbf{W}_t^{-1}\overline{\tilde{{\textbf H}}}_{\tau}}].
\end{aligned}
\label{eq:54}
\tag{54}
\end{equation}

MMSE update algorithm $\textbf F_{\text{BB,c}(t+1)}$ requires $\frac{4}{3}n^3$ FLOPs, MMSE requires $2n^2m+m$ FLOPs. Unlike the approach in \cite{b41}, which reduces complexity at the cost of significant performance degradation, our method is inspired by the recursive techniques in \cite{b30}. We derive a low-complexity update algorithm specifically for MMSE-THP by utilizing the augmented channel matrix, defined as $\Check{ \textbf H}_t^{\text{T}} =[\overline{\tilde{{\textbf H}}}_{t}^{\text{T}} \ \gamma\textbf{I}_{N_{\text{r}}}] = \Check{\textbf{Q}_t}\Check{\textbf{R}_t}$. 
\begin{equation}
\begin{aligned}
&\Check{\textbf{Q}}_t\Check{\textbf{R}}_t =  \begin{bmatrix}
\textbf Q_1   \\
\textbf Q_2 \\
\end{bmatrix}\Check{\textbf{R}}_t,\quad\overline{\tilde{\textbf{H}}} = \textbf{Q}_1\Check{\textbf{R}}_t,
\quad\Check{\textbf{H}}_{t+1} = \begin{bmatrix}
\overline{\tilde{{\textbf{H}}}}_{t} &\overline{\tilde{{\textbf H}}}_{\tau} \\
\gamma\textbf{I}_{N_{\text{r}}} &0\\
0&\gamma\textbf{I}_{\tau}
\end{bmatrix}
\end{aligned}.
\label{eq:55}
\tag{55} 
\end{equation}

Based on \cite{b41}, we find a suboptimal solution for MMSE-THP to reduce computational complexity.
\begin{equation}
\begin{aligned}
{\Check{ \textbf H}_{a}} 
  & =\begin{bmatrix}
\Check{\textbf{Q}}_t&\Check{ \textbf H}_{\tau}    \\
\end{bmatrix}\begin{bmatrix}
\Check{\textbf{R}}_t&0   \\0
&1  \\
\end{bmatrix}=\begin{bmatrix}
\Check{\textbf{Q}}_t&\textbf q     \\
\end{bmatrix}\begin{bmatrix}
\textbf I&\textbf r   \\0
& \alpha \\
\end{bmatrix}\begin{bmatrix}
\Check{\textbf{R}}_t&0   \\0
&1  \\
\end{bmatrix}  \\ &= \underbrace{[\Check{\textbf{Q}}_t \ \textbf q]}_{\Check{\textbf Q}'}\underbrace{\begin{bmatrix}
\Check{\textbf{R}}_t&\textbf r   \\0
&\alpha \\
\end{bmatrix}}_{\Check{\textbf R'}}\\
\Check{\textbf H}_{t+1}&=[\Check{\textbf H}_t\ \Check{\textbf H}_{\tau}]\textbf E^{\text{T}}=   \underbrace{\Check{\textbf Q'}\textbf G_{\text{v}}^{\text{T}}}_{\Check{\textbf Q}_{t+1}}\underbrace{\textbf G_{\text{v}} \Check{\textbf R}'\textbf E^{\text{T}}}_{\Check{\textbf R}_{t+1}},
\end{aligned}
\tag{56}
\end{equation}
where $\textbf q = \alpha^{-1}(\textbf I-\Check{\textbf Q}_{t}\Check{\textbf Q}_t^{\text{T}})\Check{\textbf H}_{\tau}$, $\textbf r = \Check{\textbf Q}_{t}^{\text{T}}\Check{\textbf H}_{\tau}^{\text{T}}$, $\alpha =\|(\textbf I - \Check{\textbf Q}_t\Check{\textbf Q}_t^{\text{T}})\|_2$.

\begin{equation}
\begin{aligned}
&\Check{\textbf{F}}_{t+1} =\Check{\textbf{Q}}'\textbf{G}_{\text{v}}^{\text{T}}\quad\Check{\textbf{R}}_{t+1} = \textbf{G}_{\text{v}}\Check{\textbf{R}}'\textbf{E}^{\text{T}}\\
&\Check{\textbf{G}}_{t+1} =  \begin{bmatrix}
\Check{\textbf{G}}_{t}(:,1:\tau)   \\
\text{diag} (\Check{\textbf{R}}_{t}(:,\tau:N_{\text{r}}+1)) \\
\end{bmatrix}\\&\Check{\textbf{B}}_{t+1} =  \begin{bmatrix}
\Check{\textbf{B}}_{t}(1:\tau-1,:)   \\
\Check{\textbf{R}}_{t}(\tau:N_\text{r}+1,:)\Check{\textbf{G}}_{t+1} \\
\end{bmatrix}.
\end{aligned}
\label{eq:57}
\tag{57} 
\end{equation}
\begin{algorithm}[t]
    \caption{RS Update Algorithm}
	\label{algorithm3} 
	\renewcommand{\algorithmicrequire}{\textbf{Input:}}
	\renewcommand{\algorithmicensure}{\textbf{Output:}}
	\begin{algorithmic}[1]
		\REQUIRE $\textbf s$, $\textbf H$ $N_{\textnormal{t}}$, $N_{\textnormal{r}}$, $\text{SINR}$, $\textbf F_{\text{BB,c}(t)}$ and $\textbf F_{\text{BB,p}(t)}$;
             
		\ENSURE $\textbf F_{\text{BB,c}(t+1)} \leftarrow $Eq.~\eqref{eq:54} and $\textbf F_{\text{BB,p}(t+1)} \leftarrow$ Eq.~ \eqref{eq:57}.
            
            \STATE  $\textbf F_{\text{BB,c}(t)}$ and $\textbf F_{\text{BB,p}(t)}$  $\leftarrow$ \eqref{eq:54},\eqref{eq:57}
            ;
          \IF{$N_{\text{u}} == 0$ and constraints \eqref{eq:44}}
     \STATE $\textbf F_{\text{BB,c}(t)}$ and $\textbf F_{\text{BB,p}(t)}$;
          \RETURN;
          \ELSE       
      \STATE $\textbf F_{\text{BB,c}(t+1)}\leftarrow $ Eq.~\eqref{eq:52},Eq.~\eqref{eq:53};
     \STATE  $\textbf F_{\text{BB,p}(t+1)} \leftarrow$
      Eq.~\eqref{eq:55}-Eq.~\eqref{eq:57};
      \RETURN;
          \ENDIF

             \RETURN  
	\end{algorithmic} 
\end{algorithm}

The proposed RS-based update algorithm is summarized in Algorithm~\ref{algorithm3}, where the sensing constraint is given by \eqref{eq:44}.

\subsection{Complexity Analysis}
Since the computational complexity remains uniform across each iteration of the AO framework, we focus our analysis on the precoding algorithm to demonstrate how the proposed update precoding effectively mitigates the overall computational complexity. We use floating point operations (FLOPs) as a computational metric to analyze and compare the complexity of various algorithms. For consistency in our analysis, the system dimensions are defined as $m=N_\text{t},n=N_{\text{r}}$, representing the number of transmit and receive antennas, respectively. 
The number of permutations for MB is equal to $M$. Let $N_{\text{u}}$ represent the variation in the number of users.


\begin{table}[t]
\begin{center}
   \caption{Complexity of Precoding Algorithm.}
   \renewcommand{\arraystretch}{2.5} 
   \begin{tabular}{c|c}  
      \hline
      \textbf{Algorithm} & \textbf{FLOPs} \\
      \hline
      \makecell{Optimization ZF } & 
        $\begin{aligned}
         & \frac{2}{3}n^3 + 2nm^2 + mn + m^2  +m + M 
        \end{aligned}$ \\
      \makecell{Optimization  MMSE } & 
        $\begin{aligned}
          \frac{2}{3}n^3 &+ 2mn^2 + 2nm^2 + mn  \\&+ m^2  +m + M
          
        \end{aligned}$ \\
      \makecell{MB RS \\ ZF-cTHP } & 
        $\begin{aligned}
         & \frac{2}{3}n^3 + 2nN_{\text{RF}}^2 + N_{\text{RF}}n + N_{\text{RF}}^2  +N_{\text{RF}} + M 
        \end{aligned}$ \\
      \makecell{MB RS \\ MMSE-cTHP } & 
        $\begin{aligned}
         \frac{2}{3}n^3 &+ 2N_{\text{RF}}n^2 + 2nm^2 + mn  \\&+ N_{\text{RF}}^2  +N_{\text{RF}} + M
          
        \end{aligned}$ \\
         \makecell{MB RS \\ ZF-cTHP U} & 
        $\begin{aligned}
         &(n+N_{\text{u}})N_{\text{RF}}+(n+N_{\text{u}})^2N_{\text{RF}} + M 
        \end{aligned}$ \\
          \makecell{MB RS \\ MMSE-cTHP U \\} & 
        $\begin{aligned}
          (n+N_{\text{u}}) &N_{\text{RF}}+(n+N_{\text{u}})^2N_{\text{RF}} \\&+N_{\text{RF}} + M
        \end{aligned}$ \\
      \hline
   \end{tabular}
 
   \label{tab:1}
\end{center}
\end{table}

The classic ZF scheme  requires $2nm^2$ FLOPs, while the ZF-THP at the CPU requires only $\frac{2}{3}n^3 + mn^2 + mn$ FLOPs. Similarly, implementing MMSE-THP at the CPU entails a complexity of $\frac{2}{3}n^3 + mn^2 + mn + m$ FLOPs. Building upon the CF hybrid precoding framework discussed above, this paper introduces MB RS to further enhance system performance. Specifically, the proposed update algorithm exhibits significantly reduced computational complexity compared to conventional approaches. This underscores the effectiveness and efficiency of our proposed algorithmic scheme. The complexity of the various precoding algorithms is summarized in Table \ref{tab:1}. The update precoding complexity is expressed in grams as o $\mathcal{O} ((n+N_{\text{u}}) ^ 2N_ {\text {RF}} )$. Consequently, the total computational complexity is $\mathcal{O}(I_{\text{SCA}}(N_{\text{RF}}(N_r + 1))^{3.5}$.

\begin{figure}[t]
\vspace{-0.28cm} 
\centerline{\includegraphics[width=0.3\textwidth]{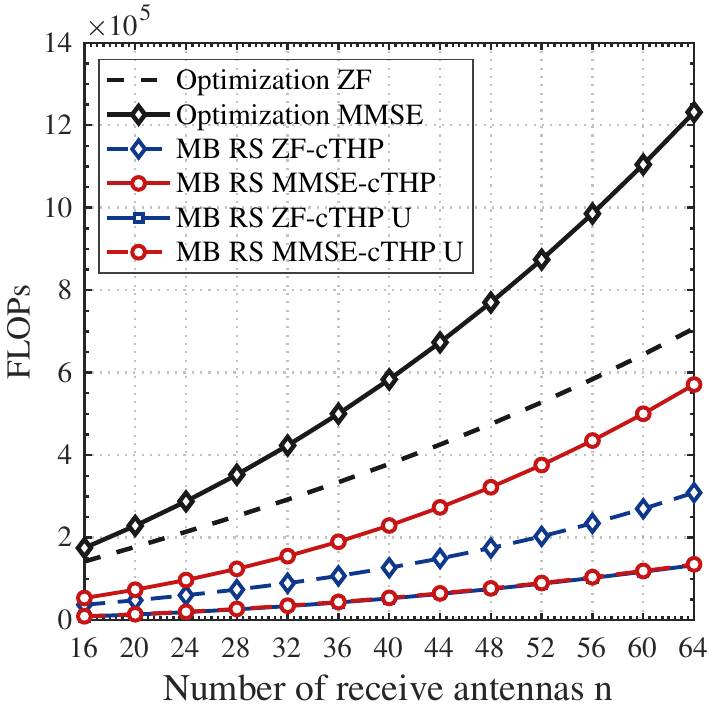}}
\caption{Complexity FLOPs of different precoding algorithms versus $n$.}
\label{fig:4}
\end{figure}

It is observed that the computational complexity escalates significantly as the number of transmit antennas increases. To address this issue, we employ an update algorithm to reduce the complexity at the CPU. For the simulation, each AP is configured with $N_{\mathrm{l}} = 64$ transmit antennas, and APs are partitioned into $L = 7$ clusters with $N_{\text{RF}} = 32$ RF chains. The simulation results in Fig. \ref{fig:4} demonstrate that the computational complexity of our proposed distributed precoding at the CPU is considerably lower than that of conventional linear precoding. Specifically, at $n = 16$, the updated MB RS ZF-cTHP scheme achieves a complexity reduction of $96.75\%$ compared to the  MB RS ZF-cTHP. Furthermore, the updated MB RS MMSE-cTHP reduces the complexity by $87.02\%$ compared to  MB RS MMSE-cTHP.

\section{SIMULATION RESULTS}
\label{V}
\begin{figure}[t]
\vspace{-0.1cm} 
\centerline{\includegraphics[width=0.31\textwidth]{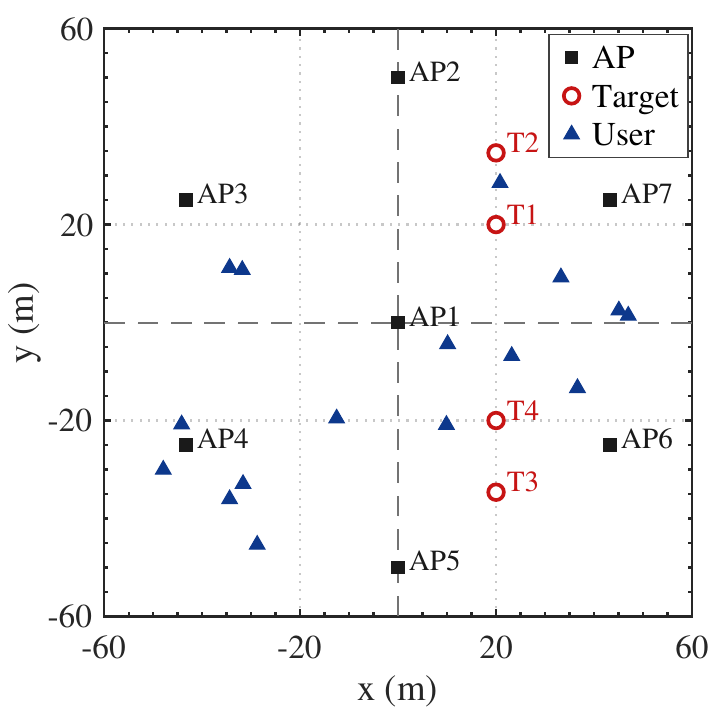 }}
\caption{Network layout and target location $\textbf{L}_{\text{o}}$
considered in this part.}
\label{fig:5}
\end{figure}
This section presents the numerical results and performance analysis of the proposed algorithms. For the simulation setup, we consider a CF massive MIMO ISAC system where each AP is equipped with $N_{\textnormal{l}}=64$  transmit antennas to broadcast
data to $k = 16$ users with $N_k =2$ receiving antennas. The user locations are randomly generated within the considered service area. The network layout of APs and target positions is shown in \figureref{fig:5}. In the considered spatial layout, the first AP1 is deployed at the origin $L_{\textnormal{AP1}} = (0, 0)$. The spatial coordinates of the four  sensing targets are configured as follows: Target 1 is located at $L_{o(1)} = (20, 20)$, Target 2 at $L_{o(2)} = (20, 20\sqrt{3})$,  Target 3 at $L_{o(3)} = (20, -20\sqrt{3})$ , and Target 4 at $L_{o(4)} = (20, -20)$. We evaluate the ESR performance of the optimized hybrid precoding for communication, and validate the beamforming gain in a multi-target sensing scenario. To evaluate the system performance under imperfect CSI, the channel estimation error variance is set to $\sigma_e^2 = 0.05$. 

\begin{figure}[t]
\vspace{-0.28cm} 
\centerline{\includegraphics[width=0.31
\textwidth]
{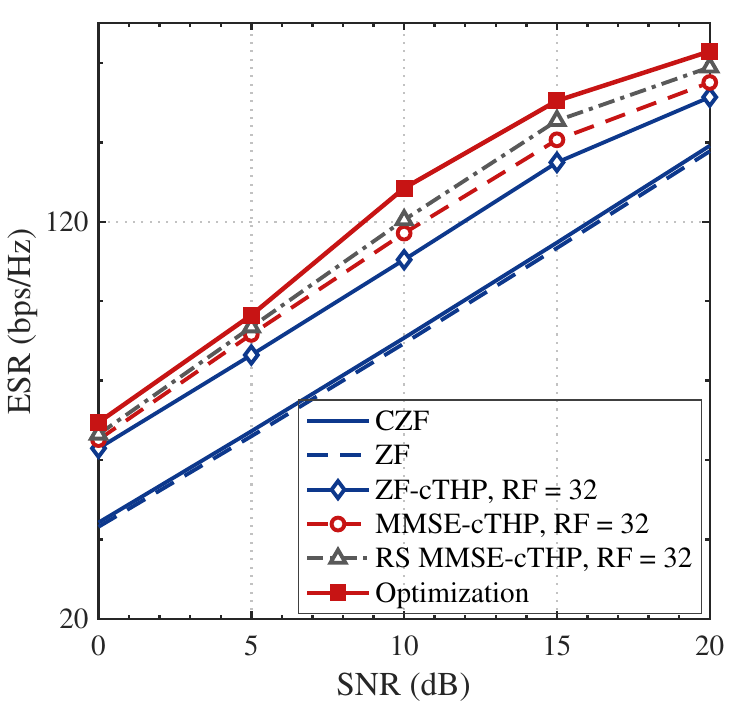}}
\caption{ESR of different precoding versus different \text{SNR}.}
\label{fig:6}
\end{figure}

\begin{figure}[t]
\vspace{0cm} 
\centerline{\includegraphics[width=0.32\textwidth]
{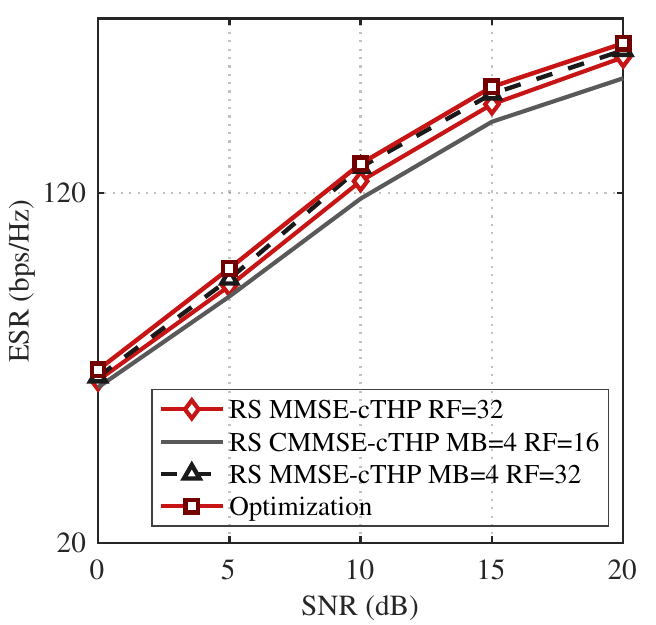}}
\caption{ESR of different MB precoding versus different \text{SNR}.}
\label{fig:7}
\end{figure}
\subsection{Communication and Sensing Analysis}

As shown in \figureref{fig:6}, the performance gap between centralized and CF distributed precoding is negligible. Furthermore, the performance penalty incurred by the AO based scheme which significantly reduces transmission overhead and computational complexity is also marginal, thereby validating the robustness of the proposed algorithm.

As illustrated in Fig.~\ref{fig:7}, we simulate and analyze the impact of the MB and partially-connected architectures on the communication performance of precoding. Specifically, this evaluation considers RF chain configurations of $N_{\text{RF}} = \{16, 32\}$, and  $\text{MB} = \{0, 4\}$. Based on the complexity analysis in Table~\ref{tab:1}, the additional computational overhead introduced by the MB is virtually negligible compared to the overall processing requirements of CF massive MIMO systems. However, it is observable that while the MB scheme effectively enhances system capacity with little computational complexity, adopting an overly sparse partially-connected architecture incurs a severe penalty in communication performance. Consequently, striking a rigorous trade-off between computational complexity and achievable system performance is imperative. 

\begin{figure}[t]
\centering
\begin{minipage}[b]{0.48\columnwidth}
    \centering
 
    \includegraphics[width=\linewidth]{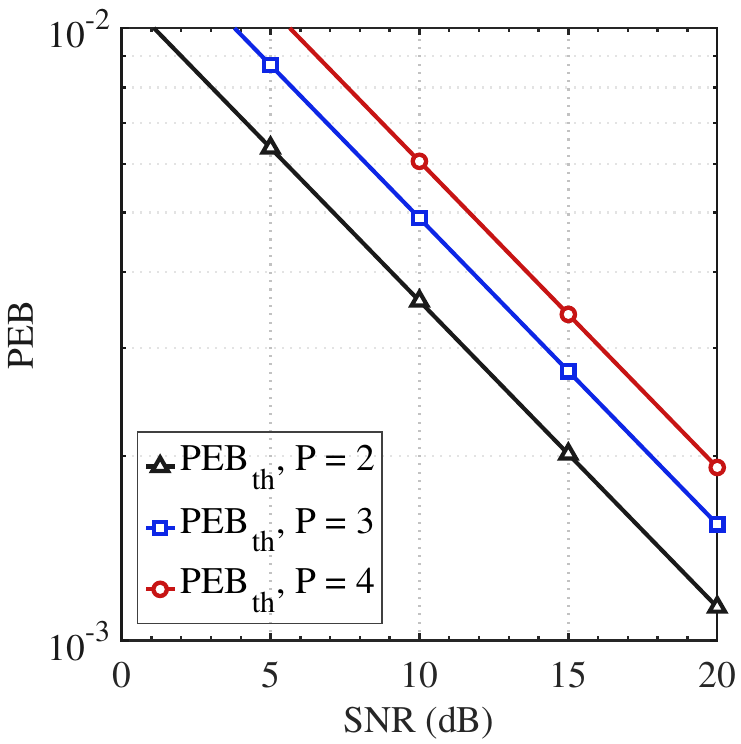}

    {\footnotesize (a) Different $\mathrm{PEB}_{\mathrm{th}}$}
\end{minipage}
\hspace{-0.03\columnwidth}
\begin{minipage}[b]{0.48\columnwidth}
    \centering
    \includegraphics[width=\linewidth]{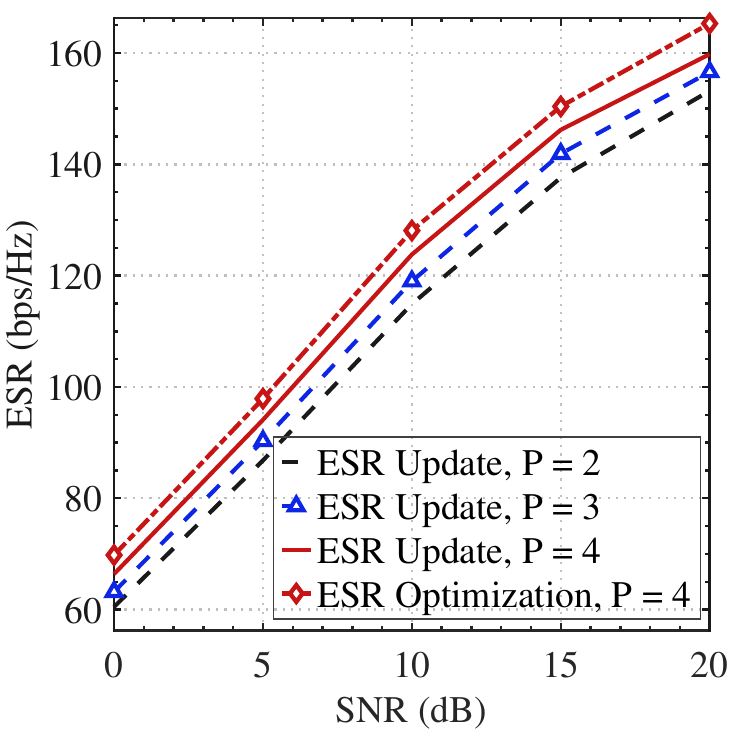}

    {\footnotesize (b) Different precoding}
\end{minipage}

\caption{C\&S performance versus different SNR.}
\label{fig:8}
\end{figure}

The sensing and communication performance is illustrated in Fig.~\ref{fig:8}. The sensing constraints are shown in Fig.~\ref{fig:8}a, and the communication performance under different PEB constraints is shown in Fig.~\ref{fig:8}b. Under identical sensing constraints, the proposed low-complexity precoding incurs practically negligible performance loss. Furthermore, as the number of sensing targets $P$ increases, the sensing environment becomes increasingly complex, exacerbating multipath propagation and mutual interference among the targets. This severe multipath effect inevitably leads to a degradation in sensing performance, as evidenced by the increased PEB. Conversely, when the system focuses on fewer targets $P=2$, it achieves highly accurate sensing. Nevertheless, such stringent sensing constraints inherently preclude the attainment of optimal communication performance. Specifically, to strictly satisfy the requisite sensing thresholds, the system is compelled to compromise its communication capacity via restricted power allocation, which ultimately leads to a deterioration of the communication ESR.

\begin{figure}[t]
\centering
\vspace{-0.6cm} 
\hspace*{-0.6cm}
\includegraphics[width=0.31\textwidth]{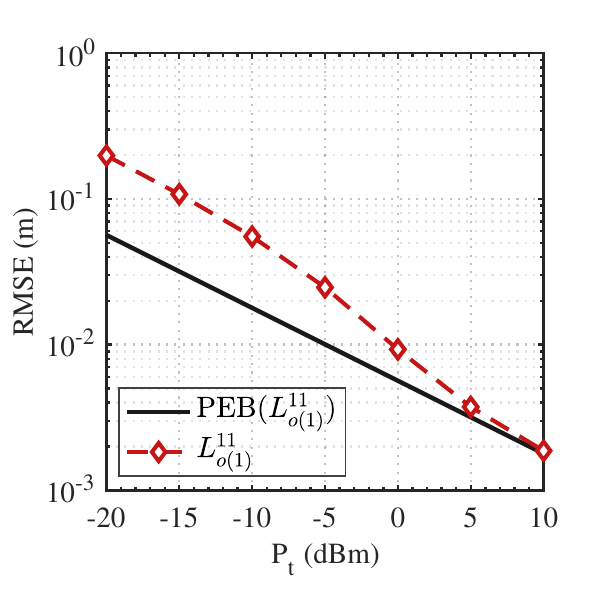}
\caption{Sensing RMSE and PEB versus different $\text{P}_{\text{t}}$.}
\label{fig:9}
\end{figure}

As illustrated in Fig. \ref{fig:9}, the RMSE performance of the position estimation is evaluated. In the high SNR , numerical results confirm that the estimator consistently converges to the global optimum, with the empirical RMSE tightly matching the theoretical PEB. Conversely, in the low SNR regime, severe random noise can overwhelm the received signal, potentially trapping the algorithm in local optima or yielding completely spurious estimates. Under such conditions, the range estimates essentially degenerate into random values uncorrelated with the ground truth, thereby introducing a pronounced performance gap between the actual RMSE and the theoretical PEB.


\begin{figure}[t]
\centering
\begin{minipage}[b]{0.48\columnwidth}
    \centering
    \includegraphics[width=\linewidth]{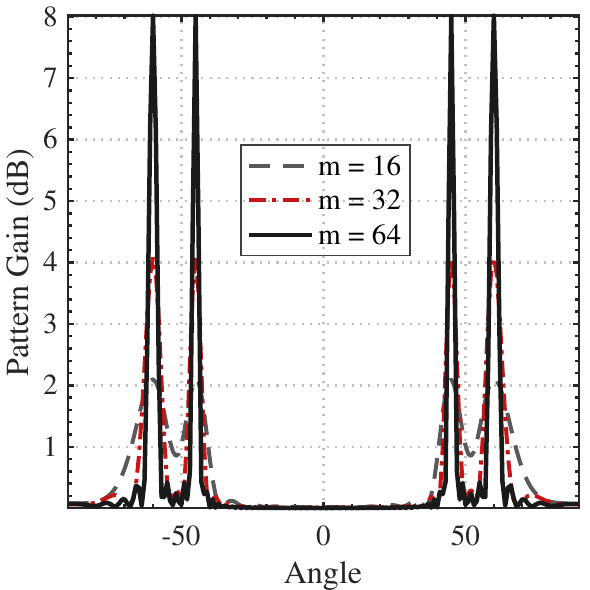}

    {\footnotesize (a) Different number of $N_{\mathrm{t}}$}
\end{minipage}
\hspace{-0.02\columnwidth}
\begin{minipage}[b]{0.48\columnwidth}
    \centering
    \includegraphics[width=\linewidth]{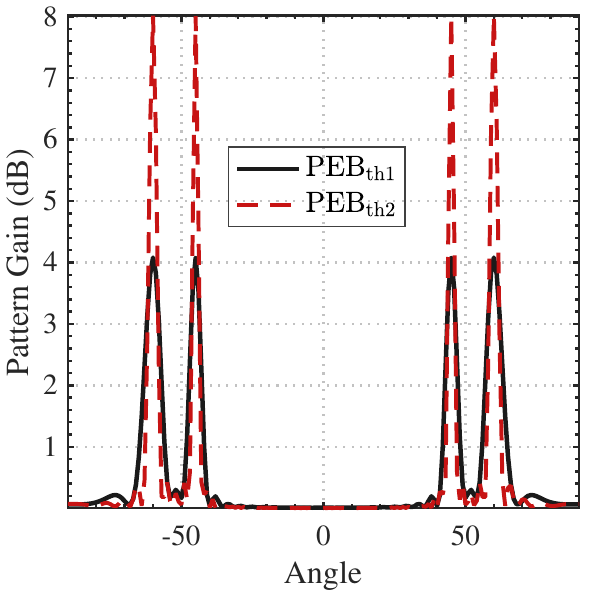}

    {\footnotesize (b) Different $\mathrm{PEB}_{\mathrm{th}}$}
\end{minipage}

\caption{The transmit beampattern versus different angle.}
\label{fig:10}
\end{figure}

\subsection{Beam Pattern}
In this subsection, we evaluate the performance of  sensing scheme in AP $l$ in terms of beampattern. For the beampattern evaluation, we focus on the transmit beampattern of $\mathrm{AP}$  $1$, which is located at the origin. 
The four target directions from  $\mathrm{AP} 1$ are approximately $\theta_1=45^\circ,\quad\theta_2=60^\circ,\quad
\theta_3=-60^\circ,\quad\theta_4=-45^\circ.$

As illustrated in  Fig.~\ref{fig:10}a,  we evaluate the impact of the transmit antenna array size, $N_{\textnormal{l}}$, on the sensing beampattern. It is evident that as $N_{\textnormal{l}}$ increases, the main lobe directed towards the target becomes progressively narrower and exhibits a substantially higher peak gain. This inversely proportional relationship between the beamwidth and the number of antennas significantly enhances both the spatial resolution and the directional beam gain of the system. However, as corroborated by the results in Fig.~\ref{fig:4}, scaling up the number of transmit antennas inevitably precipitates a surge in computational complexity. This fundamental trade-off distinctly highlights the practical necessity and superiority of our proposed low-complexity precoding update algorithm, which seamlessly maintains high beamforming gain and sharp spatial resolution while circumventing prohibitive computational complexity.

As illustrated in Fig.~\ref{fig:10}b, based on the beampattern simulations under varying sensing constraints corresponding to Fig.~\ref{fig:8}, the spatial characteristics of the generated beam are profoundly influenced by the imposed sensing constraints. Specifically, to satisfy these rigorous sensing requirements, the system is compelled to restrict its power allocation, which inherently compromises the optimal precoding design. Given that the directional beam gain is directly dictated by the precoding performance, this compromise macroscopically manifests as a conspicuous reduction in the peak beam gain. Consequently, stringent sensing constraints fundamentally bottleneck the achievable beamforming capabilities. This underscores the critical necessity of configuring an appropriate sensing threshold to strike an optimal balance between high-precision sensing and robust communication performance.

\section{CONCLUSION}
\label{VI}
This paper proposed a low-complexity hybrid precoding framework for CF massive MU-MIMO ISAC systems. By employing partially-connected RF architectures at distributed APs, the high-dimensional channel matrix is projected onto a low-dimensional channel matrix, which effectively reduces fronthaul overhead and baseband computation complexity. We formulated an ESR maximization problem with PEB constraints to jointly guarantee communication performance and multi-target sensing accuracy. To solve the resulting non-convex problem, an AO-based algorithm was proposed with convex approximation of the PEB constraint, low-dimensional digital precoding, and manifold-based analog precoding. Furthermore, an MB RS MMSE-THP update algorithm was designed to avoid repeated full matrix recomputation under dynamic user topology.
Simulation results demonstrated that the proposed framework reduced computational complexity by 87.02\%.

\bibliographystyle{IEEEtran} 
\bibliography{main}    

\end{document}